# Mathematical Harmony Analysis

*On measuring the structure, properties and consonance of harmonies, chords and melodies*

**Dr David Ryan, Edinburgh, UK**

**Table of Contents**

## 1) Abstract

Musical chords, harmonies or melodies in Just Intonation have note frequencies which are described by a base frequency multiplied by rational numbers. For any local section, these notes can be converted to some base frequency multiplied by whole positive numbers. The structure of the chord can be analysed mathematically by finding functions which are unchanged upon chord transposition. These functions are are denoted invariant, and are important for understanding the structure of harmony. Each chord described by whole numbers has a greatest common divisor, *GCD*, and a lowest common multiple, *LCM*. The ratio of these is denoted *Complexity* which is a positive whole number. The set of divisors of *Complexity* give a subset of a *p*-limit tone lattice and have both a natural ordering and a multiplicative structure. The position and orientation of the original chord, on the ordered set or on the lattice, give rise to many other invariant functions including measures for otonality and utonality. Other invariant functions can be constructed from: ratios between note pairs, prime projections, weighted chords which incorporate loudness. Given a set of conditions described by invariant functions, algorithms can be developed to find all scales or chords meeting those conditions, allowing the classification of consonant harmonies up to specified limits.

## 2) Introduction

How is it possible to make a distinction between "good harmony" and "bad harmony"? Or from a "pleasant transition" in a melody, to a "jarring transition"? These questions appear to be purely subjective, with answers dependent on the whim of a listener. However, as the ancient Greeks and Chinese discovered (using instruments like monochords) and as scientists through history have reiterated (Zarlino, Mersenne, Euler, Helmholtz) the harmonies which are more pleasing and consonant to the human ear are those where the frequency ratios are between small whole numbers.

The aim of this paper is to develop this theme further: to investigate a wide range of potential harmonies which are underutilised in modern music; to present mathematical functions which help measure the structures within harmony; to investigate the complexity of harmony and thus how consonant or dissonant it may sound. These mathematical devices and tools do not replace the subjective appreciation of a music lover; enjoyment is still a central goal. The mathematical functions are best used to supplement aesthetics, helping composers evaluate harmony, like a satnav for note choices, and provide useful tools to judge the likely impact of a note combination.

At the time of writing, the dominant tuning is equal division of the octave into twelve semitones (12-EDO), which appears to occupy a happy ground between simplicity (only 12 notes per octave on keyboard or fretboard) and complexity (it has approximate perfect fifths, major thirds and minor thirds; although the thirds are badly tuned).

Nonetheless, this paper is written regarding Just Intonation (JI). This is the original and fundamental theory of harmony discovered by the ancients, using pure harmonic intervals constructed from ratios of small whole numbers. Other tuning systems (such as equal temperaments, well temperaments or meantone) typically aim to approximate the JI intervals well (if the tuning system was intended to produce harmony at all). So to understand the structure of harmony, JI is the system to study. Moreover, it is well suited to mathematical study since its chords and harmonies are constructed from whole numbers.

What happens when the spectrum of sound available in Just Intonation is reduced to the 12 notes in Equal Temperament? A lot of good harmony is lost, and that which remains has an element of discord. Recent literature has shown a trend away from considering the whole number ratios which give intervals their consonance, and consider only what the 12-EDO note combinations are and how they progress. This is a problem since 12-EDO has no explanatory power of harmony, it only works since it approximates a small part of JI well. The structure of JI is where the explanatory power resides. 12-EDO cannot explain why a major triad should be supplied by notes ($n$, $n+4$, $n+7$) of the piano keyboard scale; why not ($n$, $n+4$, $n+8$)? The only satisfactory explanation is that the 12-EDO note pattern ($n$, $n+4$, $n+7$) approximates the JI frequency ratio 4:5:6 well. 12-EDO also cannot explain why some of the notes 'in-between' the piano keys work well (e.g. the 'barbershop' seventh, or some 'jazz' or 'blue' notes), but JI can explain this by compound frequency ratios between the notes (e.g. 4:5:6:7 for the barbershop seventh). This demonstrates the premise that *theories of musical harmony make most sense when expressed in Just Intonation*. Moreover, once theories are given in JI, it is most likely possible to translate them back into other systems of interest, e.g. to explain the above chord formation in 12-EDO. For example, the JI Tonnetz (defined below) when wrapped gives the 12-EDO Tonnetz, and it is possible to translate facts about the structure of the JI Tonnetz into similar facts about the wrapped Tonnetz.

On the subject of sevenths, and of elevenths and thirteenths, modern harmony suffers a paucity of variety and of new note combinations due to artificially restricting the scale to twelve notes. We have run out of new chords from twelve notes; witness the dramatic slide from classical harmony to atonality in a few short equal-tempered decades. Is twelve note serialism supposed to be an improvement? Unbridled chromaticism is what happens when composers run out of interesting new things to do, when all the meat has been picked off the twelve-note carcass. In comparison, Just Intonation is a wide open expanse with uncharted territories of harmony to explore. Mathematical functions help us to chart and map the more consonant harmonies in order to recognise them, tame them, make them fit, order them for compositional use. The study of these whole-numbered frequency ratios is the basis of harmony, as the historical record shows.

## 3) Literature review with commentary

Many accounts have been written regarding the history of musical tuning and harmony: the reader is referred to Fauvel, Flood & Wilson (2006) for a history of tuning and temperament; to Partch (1974) for a history of tuning with Just Intonation in mind; see also Haluska (2004) and Sethares (2005). For discussion of some of the more mathematical aspects of music, see Wright (2009). Significant historical names include Pythagoras, Zarlino, Benedetti, Mersenne, Rameau, Euler, Hauptmann, Helmholtz and Hugo Riemann; however scientists of all ages have been intrigued by the properties of musical harmony and many others have written on the subject.

Although the further back the historical record is traced, the less precise are the details, the consensus is that stringed instruments provided the main route for discovering the properties of harmony. The musical instrument called a 'monochord' is a resonant body with a single string and a moveable bridge dividing the string in two. Two notes can be produced from one string at constant tension; the lengths of the two strings can be measured precisely.

The main discovery was that the two notes sounded better together (consonant) when the ratios of the string lengths were small whole numbers, such as octaves (2:1 string length ratio) and perfect fifths (3:2). Other string lengths sounded bad (dissonant) and it was noted the ratios of string lengths were not small whole numbers; the ancient Greeks appear to have been aware of 729:512, a diminished fifth (produced via six perfect fifths) which is roughly six semitones on the modern piano; then and now this interval has always been regarded as dissonant. The large whole numbers provide an explanation for the dissonance and negative sensation produced by the diminished fifth interval.

Later it was discovered that string length was inversely proportional to frequency. This meant consonant string lengths would give consonant frequencies, and vice versa; small whole number length ratios would give small whole number frequency ratios, and vice versa. Here then is **the fundamental theorem of harmony**: that *two notes played together sound good (consonant) when their frequency ratio uses small whole numbers*; an increasingly unpleasant (dissonant) sound is produced as the numbers become larger.

In ancient Greek times it appears that frequency (or string length) ratios regarded as consonant were restricted to ratios between the numbers 1, 2, 3 and 4. These numbers yielded the octave (2:1), perfect fifth (3:2) and perfect fourth (4:3). Due to the Pythagorean tuning, produced from a cycle of perfect fifths, the major third was regarded as dissonant because four perfect fifths minus two octaves give the interval 81:64, which does not have particularly small whole numbers.

By the time of Mersenne the 'musical numbers' had expanded to be 1, 2, 3, 4, 5 and 6. This gave a major third of 5:4 and a minor third of 6:5. The major third was now fixed since 5:4 = 80:64 is subtly different (and more consonant) than the Pythagorean 81:64. Also, the major sixth of 5:3 and major tenth of 5:2 could be produced. Moreover, the three-note major triad chord (4:5:6) sounded so good it was regarded by Zarlino as being the basis of harmony, and a C major scale was built from it (1/1, 9/8, 5/4, 4/3, 3/2, 5/3, 15/8, 2/1) corresponding to the pitch classes C, D, E, F, G, A, B, with C repeated an octave up. See also Rameau's Treatise on Harmony (1722).

The genius polymath Euler, who produced a scientific work on musical harmony called the 'Tentamen' (Euler 1739), conjectured that the extension of the major triad (4:5:6) to a four-note dominant seventh chord (4:5:6:7) would produce a more complete harmony, however this chord has not (to date) overtaken the major triad in popularity. Thus music (almost) learned how to count to 7 back in the eighteenth century (Monzo 2016); the 12-EDO piano keyboard has prevented $7^{th}$ harmonics from becoming commonplace, although it could be argued that barbershop and blues music have found ways around that limitation by using instruments less limited by design considerations, such as the human voice, or the bent guitar string.

The general theme therefore is that as time progressed, the numbers regarded as musical have expanded through 1..4, 1..6, 1..7, which are respectively 3-limit, 5-limit and 7-limit tuning. Were modernity to run out of new musical chords, 11-limit and 13-limit tunings would be good places to start looking for new chords; in fact, free choice of notes in JI (no-limit tuning!) might be even more useful, where small prime numbers (say up to 60 or 128) could be used freely for varying aesthetic effects and varying levels of consonance relating to their prime height.

So much for the musical numbers. These have been described in detail to show that the basis of consonant harmony is no mystery, but entirely predictable from the fundamental theorem of harmony, of frequency ratios using small numbers giving better harmonies.

The question is how to quantify this consonance? What tools have previous authors developed in order to separate out better from worse harmonies? The paragraphs which follow describe some existing functions from literature.

$$BH\left(\frac{a}{b}\right) = BenedettiHeight\left(\frac{a}{b}\right) = a \cdot b \quad \text{(where } \frac{a}{b} \text{ have no common factors)}$$ **Equation 1**

(In this paper the convention is that most functions have both a descriptive name, e.g. *BenedettiHeight*, and an abbreviated name, e.g. *BH*, in order to aid both textual description of music and concise mathematical terminology. Both forms will be used interchangeably.)

In Equation 1 the *BenedettiHeight* function (attributed to Giovanni Benedetti, see Drake 1970, Monzo 2016, Xenharmonic 2016) takes a musical frequency ratio *a*:*b* (also written as *a*/*b*) and returns *a* multiplied by *b*. For this function it is important that all common factors have been divided out of *a* and *b*, i.e. that they are in lowest terms, that they are coprime. Example values for some simple intervals include: *BH*(2/1) = 2, *BH*(3/2) = 6, *BH*(5/4) = 20, *BH*(9/8) = 72 for the octave, perfect fifth, major third and major whole tone respectively.

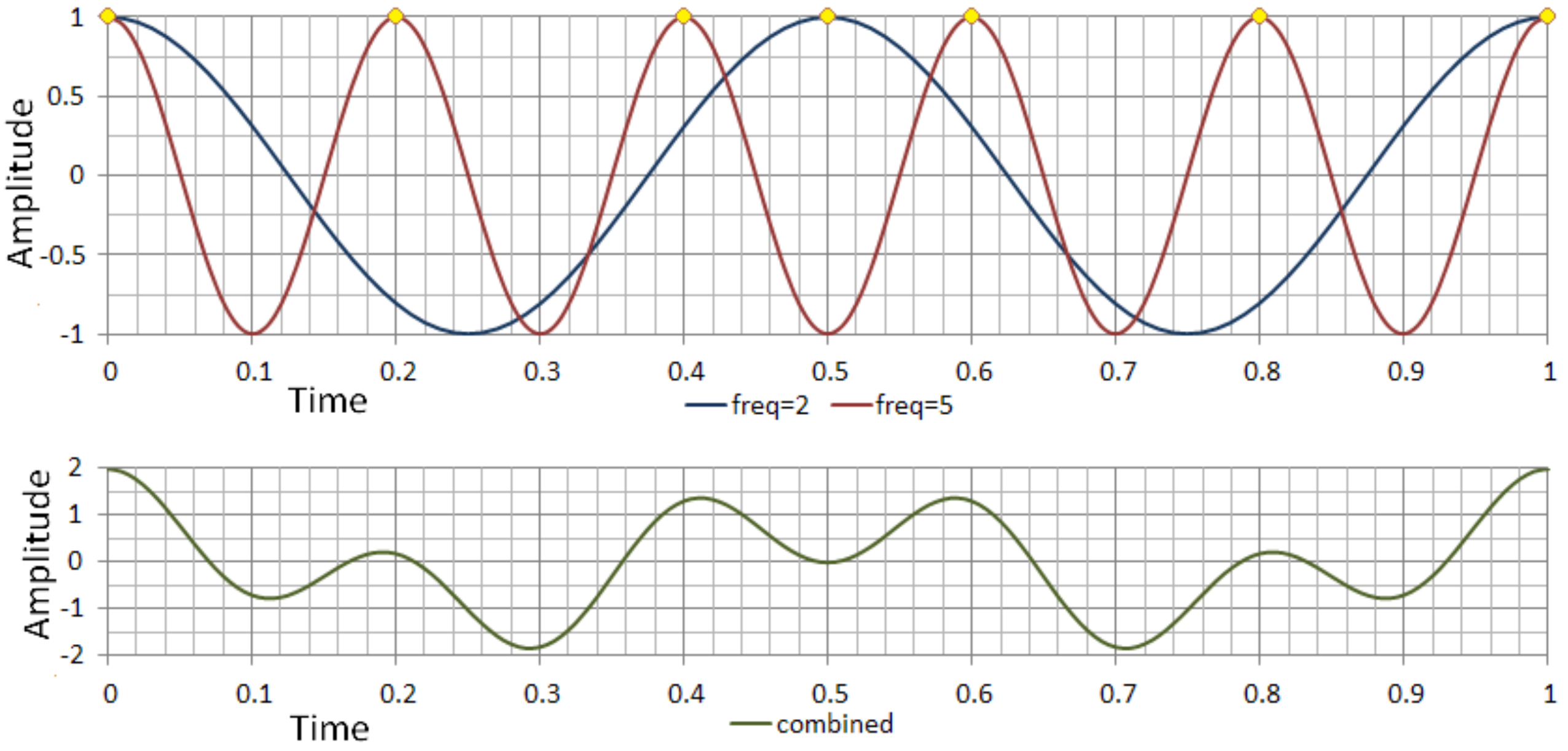

**Figure 1: a) Graph of amplitude vs time for two sine waves of frequency 2 and 5.**
**Peak values are highlighted with yellow dots.**
**b) The combined waveform with ratio 10 between largest and smallest feature scales**

What exactly is being measured by the Benedetti height? In Figure 1a two sine waves are plotted which both have maximum amplitude at time zero. Each time either waveform reaches maximum amplitude, a yellow dot is added. All the yellow dots are at multiples of time 0.1 = (1/10). In general, with coprime frequencies $a$ and $b$, the peak amplitudes would be reached at times a multiple of $(a \cdot b)^{-1} = (BH)^{-1}$, so this is the minimum feature size. Also, the whole waveform repeats with period 1 in Figure 1b, so 1 is the maximum feature size. Hence the number *BH* is the ratio of minimum to maximum feature sizes of the combined waveform, and the Benedetti height could be described as a ‘complexity’ measure for adding these two sine waves.

For *BenedettiHeight* a lower number means a more consonant interval, and a higher number means a more dissonant interval. In this way, consonance and dissonance become a sliding scale, with no objective cut-off point between the two. Were a particular cut-off point artificially introduced, however, only a finite number of intervals would be consonant with respect to that cut-off point, and an infinite number would remain dissonant. For example, up to *BH* = 6 the only (reduced) ratios possible are 2:1, 3:1, 3:2, 4:1, 5:1 and 6:1. Hence there are always a finite number of consonant intervals but an infinite number of dissonant intervals. This explains why musical harmony has always focused on a small number of consonant intervals, and it explains why the consonant intervals are special and exceptional cases in terms of their aural and aesthetic qualities.

$$KH\left(\frac{a}{b}\right) = KeesHeight\left(\frac{a}{b}\right) = \max(a', b') \quad (a', b' \text{the odd parts of } a, b \text{ which are coprime})$$ **Equation 2**

In Equation 2 (Xenharmonic 2016) the Kees height is defined. It will be demonstrated for the frequency ratio 28:30. Firstly the lowest terms are found, to get rid of any common factors and make the ratio coprime: 28:30 –> 14:15. Secondly the odd parts are found, which means dividing each number by 2 until it becomes an odd number (e.g. discarding the power of 2 in the prime factorisation), so 14:15 becomes 7:15 which is a ratio between odd numbers. Finally, find the maximum of these: hence *KH*(28:30) = 15.

This function *KeesHeight* can be used to define the '$q$ odd limit' intervals which are all the intervals using only odd numbers up to $q$ (after discarding all powers of 2 and common factors). Compare this with prime limit '$p$-limit' tunings which allow any odd numbers as long as no prime factor is above $p$. Prime limits are more normally used for tunings since they allow composite odd numbers to appear much earlier, for example 15:8 with a Kees height of 15 appears first in odd limit 15, but prime limit 5. Since 15:8 is consonant with both 3:2 and 5:4 it makes sense for it to appear at the same time as them, and for 15:8 to appear earlier than intervals like 11:8 and 13:8 which have lower Kees height. Hence there are consonance benefits of using prime limits rather than odd limits. However, if the Kees height was used, higher values also indicate lower consonance. Kees height would make most sense to use on an instrument such as a diamond marimba where the *KH* value is a primary design consideration.

$$TH\left(\frac{a}{b}\right) = TenneyHeight = \log_2\left[BH\left(\frac{a}{b}\right)\right] \quad (a, b \text{ coprime}) \qquad \textbf{Equation 3}$$

In Equation 3 (Xenharmonic 2016) the Tenney height is given. This is the base-2 logarithm of the Benedetti height. An advantage of *TH* is that intervals of $n$ octaves ($2^n$:1) have Benedetti height of $2^n$, but Tenney height of $n$, so the Tenney height uses smaller numbers. For ratios using numbers up to 100, *BH* takes values up to 10000, but *TH* only up to 13.29. Hence the size of *TH* is far more convenient, at the expense of usually being an irrational decimal number. *BH* remains an integer, and also retains the prime factorisation of the original chord which is useful for some applications.

Euler (1739) extended the 2-note *BenedettiHeight* function to an $n$-note function we shall call the Euler sweetness function (*ESF*), also known as the Euler softness function (Monzo 2016). For a compound ratio (e.g. 4:6:8) the greatest common divisor (*GCD*) is found, and then divided out of each number in the ratio (e.g. *GCD*(4:6:8) = 2, so the ratio becomes 2:3:4). Then the lowest common multiple is found of the reduced ratio (e.g. *LCM*(2:3:4) = 12 = $2^2$ 3). Then finally *ESF* is the sum of 1 and ($p$-1) for each prime $p$ dividing into the *LCM*, where each prime can be counted multiple times (e.g. 12 = 2×2×3, so *ESF*(4:6:8) = 1+(2-1)+(2-1)+(3-1) = 5).

| | |
|---|---|
| I | 1; |
| II | 2; |
| III | 3, 4; |
| IV | 6, 8; |
| V | 5, 9, 12, 16; |
| VI | 10, 18, 24, 32. |
| VII | 7, 15, 20, 27, 36, 48, 64; |
| VIII | 14, 30, 40, 54, 72, 96, 128; |
| IX | 21, 25, 28, 45, 60, 80, 81, 108, 144, 192, 256; |
| X | 42, 50, 56, 90, 120, 160, 162, 216, 288, 384, 512; |
| XI | 11, 35, 63, 75, 84, 100, 112, 135, 180, 240, 243, 320, 324, 432, 576, 768, 1024; |
| XII | 22, 70, 126, 150, 168, 200, 224, 270, 360, 480, 486, 640, 648, 864, 1152, 1536, 2048; |
| XIII | 13, 33, 44, 49, 105, 125, 140, 189, 225, 252, 300, 336, 400, 405, 448, 540, 720, 729, 960, 972, 1280, 1296, 1728, 2304, 3072, 4096; |
| XIV | 26, 66, 88, 98, 210, 250, 280, 378, 450, 504, 600, 672, 800, 810, 896, 1080, 1440, 1458, 1920, 1944, 2560, 2592, 3456, 4608, 6144, 8192; |
| XV | 39, 52, 55, 99, 132, 147, 175, 176, 196, 315, 375, 450, 500, 560, 567, 675, 756, 900, 1008, 1200, 1215, 1344, 1600, 1620, 1792, 2160, 2187, 2880, 2916, 3840, 3888, 5120, 5184, 6912, 9216, 12288, 16384; |
| XVI | 78, 104, 110, 198, 264, 294, 350, 352, 392, 630, 750, 840, 1000, 1120, 1134, 1350, 1512, 1800, 2016, 2400, 2430, 2688, 3200, 3240, 3584, 4320, 4374, 5760, 5832, 7680, 7776, 10240, 10368, 13824, 18432, 24576, 32768. |

**Figure 2: Euler's degrees of sweetness / softness (reproduced from Monzo 2016)**

*ESF* assigns every 2-note interval and every *n*-note chord a 'degree', a positive whole number by which the different intervals or chords can be grouped. In Figure 2 the *ESF* degree is on the left, and the possible *LCM* values are on the right. One interesting feature of Euler's function is that when the *LCM* doubles, the *ESF* increases by 1; when the *LCM* triples, the *ESF* increases by 2. Another feature is that if *LCM* is prime, then *ESF* = *LCM*. One important feature is that adding a note to a chord will multiply *LCM* by an integer (which might be 1), so *ESF* will either stay constant or increase according to the integer. Hence *ESF* is non-decreasing when notes are added to chords.

| | |
|---|---|
| II : | **1:2** |
| III : | 1:3, 1:4 |
| IV : | 1:6, **2:3**, 1:8 |
| V : | 1:5, 1:9, 1:12, **3:4**, 1:16 |
| VI : | 1:10, 2:5, 1:18, 2:9, 1:24, 3:8, 1:32 |
| VII : | 1:7, 1:15, **3:5**, 1:20, **4:5**, 1:27, 1:36, 4:9, 1:48, 3:16; 1:64 |
| VIII : | 1:14, 2:7, 1:30, 2:15, 3:10, **5:6**, 1:40, **5:8**, 1:54, 2:27, 1:72, 8:9, 1:96, 3:32, 1:128 |
| IX : | 1:21, 3:7, 1:25, 1:28, 4:7, 1:45, 5:9, 1:60, 3:20, 4:15, 5:12, 1:80, 5:16, 1:81, 1:108, 4:27, 1:144, 9:16, 1:192, 3:64, 1:256 |
| X : | 1:42, 3:14, 6:7, 1:50, 2:25, 1:56, 7:8, 1:90, 2:45, 5:18, 9:10, 1:120, 3:40, 5:24, 8:15, 1:160, 5:32, 1:162, 2:81, 1:216, 8:27, 1:288, 9:32, 1:384, 3:128, 1:512 |

**Figure 3: Euler's classification of ratios – in bold the traditionally consonant intervals in the octave, and underlined the traditionally dissonant intervals (reproduced from Monzo 2016)**

| | | |
|---|---|---|
| II | octave | 1:2 |
| IV | quinte | 2:3 |
| V | quarte | 3:4 |
| VII | 6e maj. | 3:5 |
| VII | 3e maj. | 4:5 |
| VIII | 3e min. | 5:6 |
| VIII | 6e min. | 5:8 |
| VIII | ton maj. | 8:9 |
| IX | gde 7e min | 5:9 |
| IX | pte 7e min. | 9:16 |
| X | ton min. | 9:10 |
| X | 7e maj. | 8:15 |

**Figure 4: Euler's classification of intervals in the octave. Upper group are the consonant intervals, lower group are the dissonant intervals (reproduced from Monzo 2016)**

For two note frequency ratios, *ESF* has been evaluated up to *ESF* = 10 in Figure 3. The traditional classification of intervals in the octave (Monzo 2016) have been given in Figure 4, and highlighted in Figure 3 using bold for the consonant intervals, and underlining for the dissonant intervals. This evaluation shows that *ESF* is larger for more dissonant intervals. Overall *ESF* is a useful device for measuring the increased dissonance as the whole numbers in ratios increase, however it does not retain the information about prime numbers. Euler's intermediate step of using the *LCM* function **does** however retain prime information, so the *LCM* function will be used in what follows.

Euler also defined the concept of a 'complete' chord (Monzo 2016), which was essentially one in which no extra note could be added in without increasing the *ESF* (and the intermediate *LCM*). As an example, the chord 1:2:3 has *GCD* = 1 (so no need to reduce the ratio) and *LCM* = 6. So this chord is not 'complete', since the divisors of *LCM* are (1, 2, 3, 6) and the frequency 6 can be added in without increasing *LCM* or *ESF*. But 1:2:3:6 **is** a complete chord, since no extra number divides the *LCM* value; adding in any other whole number would increase the *LCM* by an integer factor, and in turn increase *ESF*. This concept of completeness will be developed later in terms of divisor lattices.

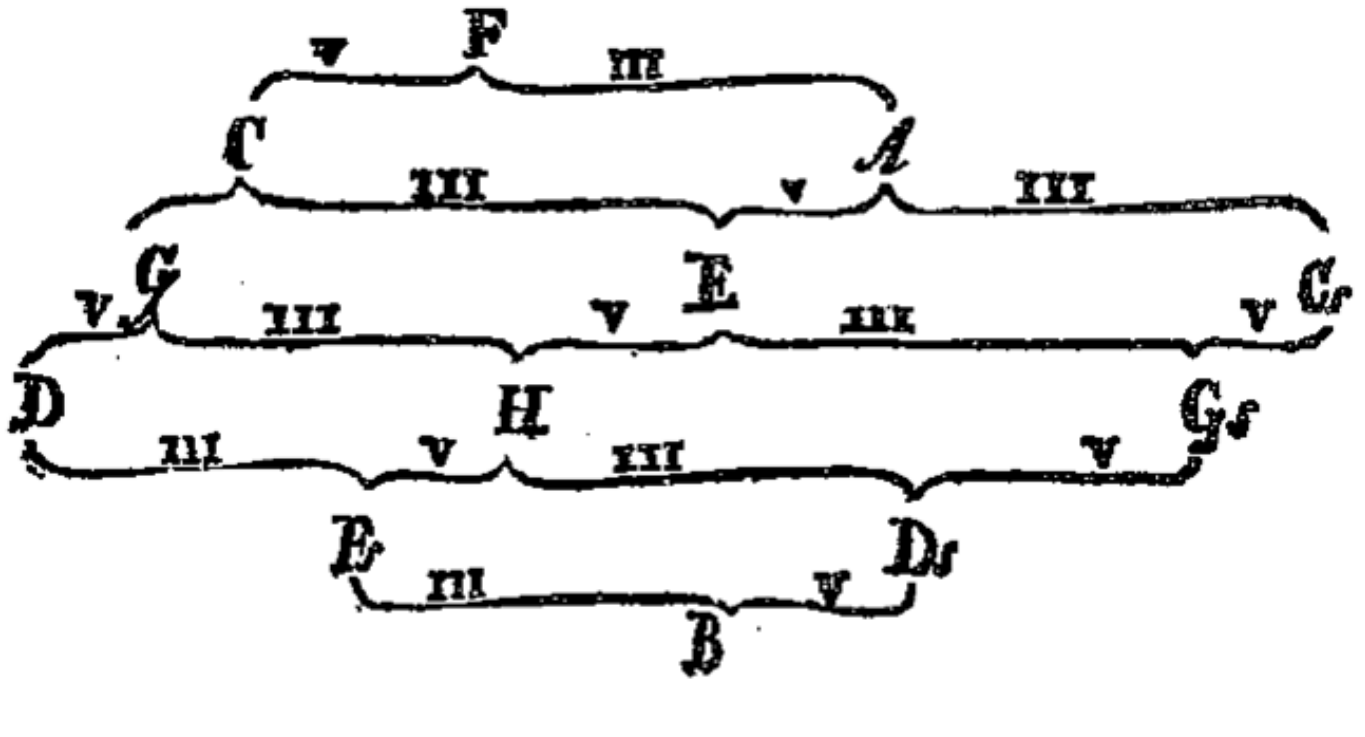

**Figure 5: Euler's tone lattice (reproduced from the Tentamen, 1739)**

Another device to be brought out of literature is the concept of 'Tonnetz', or 'tone lattice'. This, again, appears to have originated with Euler (1739) and in Figure 5 his diagram is given linking pitch classes by perfect fifths in one direction (V in Roman numerals, linking F, C, G, D), and by major thirds in another direction (III, linking C, E, G#). The tone lattice is therefore a structure in 2 or more dimensions which links pitch classes (notes up to octave equivalence) by the interval needed to travel between them. Each direction corresponds to a different prime number: a perfect fifth is a translation or movement by prime 3 (octave-equivalent to 3/2), and a major third is a translation by 5 (or 5/4). Hence Euler's diagram is for a 3-5 Tonnetz.

By adding in an extra direction for octave transformations, a 2-3-5 Tonnetz in three dimensions can be created linking individual pitches, instead of pitch classes. Extra dimensions can also be added for primes 7, 11, 13 etc, making the Tonnetz an extremely powerful descriptive device. Chords in $p$-limit tuning correspond to sets of lattice points on an $n$-dimensional Tonnetz, where $p$ is the $n$th prime number. Transposition of $p$-limit chords corresponds to movement on the $n$-dimensional Tonnetz. Sets of pitch classes (which discard information about the prime 2) correspond to an ($n$-1)-dimensional Tonnetz. Therefore, the Tonnetz is the natural 'geometry' of harmony and gives every chord a shape. The Tonnetz provides the fundamental link between the musical subject of harmony and the mathematical subject of geometry. This correspondence can be mined extensively for musical insight. Understanding the Tonnetz means understanding the structure of harmony.

Since Euler, others have rediscovered or disseminated the concept of Tonnetz, see Naumann (1858), von Oettingen (1866), Riemann (Rehding 2003). More recent authors such as Lewin (1982) and Cohn (1997, 1998) have developed 'Neo-Riemannian theory' based around transformations on this lattice which fix two out of three notes of a specific triad, moving the final note to an adjacent pitch. These theories can account for key transposition without requiring a fixed key signature. Tonnetz works equally well in Just Intonation where each direction extends infinitely, and for equal tempered systems where each direction 'wraps around' after a finite distance and there are a finite number of pitch classes in total. However it could be said that the finite versions of Tonnetz for EDOs only give approximately consonant harmony because they correspond to wrapped versions of JI tone lattices for small primes, leaving the explanation of why certain harmonies are consonant still very much with Just Intonation.

The final concepts to be brought out of literature are those of otonality (overtone-ness) and utonality (undertone-ness). The words 'otonal' and 'utonal' originated with Partch (1974) to describe the harmonic (overtone) series and subharmonic (undertone) series respectively, phenomena known for centuries before. The overtones of a note of frequency $f$ are the frequencies $2f$, $3f$, $4f$… and the undertones are the frequencies $f/2$, $f/3$, $f/4$… Modern methods (such as electronic Fourier analysis of audio signals) have shown that a plucked string can produce all of the overtones, but will not produce the undertones since the string cannot vibrate at lower frequencies. A less technological way to demonstrate the overtones is to press a finger lightly on a string before plucking, separating the string lengths into whole number ratios (1:1 for the first overtone, the octave, 1:2 for the second overtone, the perfect twelfth, etc) which can make a guitar string into a kind of monochord.

For note pairs, they are equally otonal and utonal since the ratio $a$:$b$ has both equivalent otonal ($a$/1):($b$/1) and utonal ($ab$/$b$):($ab$/$a$) forms. However for compound ratios, chords with more than 2 notes, they are generally either more otonal or more utonal. Otonal therefore means 'better described by overtones than undertones', and utonal means 'better described by undertones than overtones'. By the word 'better', read 'smaller whole number'. The major chord 4:5:6 is the most well known otonal chord, and the minor chord 10:12:15 = (60/6):(60/5):(60/4) is utonal. So in a sense, otonal and utonal are opposites of each other, and are extensions of the concepts of 'major' and 'minor' to a much wider range of chords.

One interesting fact is that otonal and utonal versions of chords sound different to the human ear (e.g. major and minor sound different) although mathematically they are completely dual and that facts about otonal chords always give corresponding facts about utonal chords. Hence an easy way to produce new music is to invert the frequencies $f \rightarrow C/f$, for some constant C, which will map between otonal and utonal versions of a harmony, and give twice the amount of interesting harmony at the same consonance level, for no extra cost.

Tracing the otonal and utonal concepts back through history, mainly through the lens of major and minor chords; Mersenne was invigorated by the musical numbers 1 to 6, which led to major chords and otonal harmony. Helmholtz (1885) believed that minor triads were inferior in harmony to major triads, whereas Hauptmann and Riemann (see Rehding 2003) believed them to be just as harmonious as each other, the minor chords and major chords being duals of each other and being converted into each other via Riemann's transformations of the Tonnetz. On this point, application of Euler's sweetness function (*ESF*) to both 4:5:6 and 10:12:15 gives the same result, the *LCM* in each case being 60. May the intellectual debate between 'Major > Minor' and 'Major ≈ Minor' rage a long time, producing many good musical works along the way!

In summary, useful facts from literature include: the 'fundamental theorem of harmony' being that small whole number frequency ratios sound pleasant and consonant; the *BenedettiHeight* function being an (inverse) measure of consonance for two notes; Euler's 'sweetness function' (*ESF*) extending consonance measures to three or more notes; 'complete' chords for which extra notes cannot be added without increasing *ESF*; the *LCM* function which does the same but retains information about primes; the geometry of the Tonnetz with a different direction for each prime; the extension of major/minor to otonal/utonal and how they are dual concepts. These will provide the basis for development of invariant functions which describe the structure of harmony.

## 4) Invariant functions of chords

The definition of 'invariant' functions are those which are unchanged by multiplying a chord by a constant factor. This is important since it represents key transposition of chords, and the structure of a harmony should be independent of which key it is played in. For a list of invariant functions in this paper, see Appendix 1.

In Just Intonation a chord can be described by a base frequency (or reference frequency) and a set of rational numbers. For example, the chord formed from 440 Hz (Concert A), 550 Hz and 660 Hz can be described as a base frequency of 440 Hz with the rational numbers (1/1, 5/4, 3/2). The base frequency is of little interest since it can be chosen arbitrarily; changing it is no more than changing the key signature, which doesn't affect musical structure. The structure of the rational number set is what determines the structure of the harmony.

Another description of the same chord could be 220 Hz with the rational numbers (2/1, 5/2, 3/1). An important representation is when the rational number set are multiplied by a number to remove the denominators and yield whole numbers only; this would be 110 Hz with (4, 5, 6). Yet another representation could be 55 Hz with (8, 10, 12), however there is a common factor (GCD) of 2 in these whole numbers, which ought to be removed before analysing. Hence (4, 5, 6) is the important whole-number description of the chord which will yield structure information about the chord. The representation 4:5:6 is equivalent to (4, 5, 6) as used above; the two forms will be used interchangeably.

Here then are some non-invariant and some invariant aspects of the descriptions above:

**Not invariant**: choice of base frequency (110 Hz, 220 Hz, etc), choice of first note (1/1, 2/1, 4, 8, etc), choice of how to represent ratio between first and second notes as a ratio (4:5, 8:10, etc).

**Invariant**: frequency ratio between first and second notes *in the lowest terms* (5/4), the same for notes two and three (6/5); lowest terms means *dividing out by common factors*, which are given by the *GCD* function.

Note that the invariant function ‘frequency ratio’ was constructed from things which were not invariant. Notice also that the base frequencies and chord notes were allowed to move about, but the multiplicative distance in-between chord notes (the ratios) were the same each time, fixed across the multiple representations. This is the general theme of invariant functions: they are about measuring *the right kind of distances* or *fixed structures* within a chord, which do not change under musical transposition, when the base frequency is changed or the whole numbers in the chord are multiplied by a constant.

In JI the notes are described by a base frequency and a finite set of rational numbers. It is always possible to change the base frequency to obtain the notes in terms of whole numbers. Hence this paper will focus on whole number sets instead of rational number sets. Much of the terminology could also be defined for rational numbers. In particular, *GCD* and *LCM* functions could be re-defined over rational numbers to enable skipping the stage of converting rational numbers to whole numbers.

Note that the analysis of melody and harmony have much in common. Both are the analysis of sets of whole numbered frequencies – whether played at the same time, or at different times. For simplicity it is assumed below that the harmony of a single chord is being analysed, but the invariants would be similar for melody. An interesting subject for further work would be to take a reasonably long melody and plot invariant functions at various points in time $t$, and over $m$ local notes in the melody. This gives invariant functions (such as *Complexity* defined below) as two dimensional functions of $t$ and $m$.

## 5) The *Complexity* of a *Chord*

Now for some mathematical definitions. First, suppose that the chord to be analysed has $N$ distinct notes in it ($N \geq 1$), and these are described by $N$ positive whole numbers in ascending order of frequency. Let $CH(n)$ = $Chord(n)$ be the $n$th number, and let $CH$ = *Chord* be the set of all these notes, in ascending order. (The convention is that each function has both an abbreviated name and a full name.)

To describe the (invariant) complexity of a particular harmonic chord we will need the greatest common divisor *GCD* and the lowest common multiple *LCM*. *GCD* is the largest number which divides a whole number of times into every frequency in *Chord*; *LCM* is the smallest number into which every frequency of *Chord* divides a whole number of times. If the *Chord* ratio is reduced (e.g. 4:5:6) then *GCD* will be 1;

if the ratio is not reduced (e.g. 12:15:18) then *GCD* will be greater than 1. In every case, the *GCD* is smaller than (or equal to) the notes in *Chord*. However the *LCM* is always larger than all the numbers in the chord; the *LCM* of 4:5:6 is 60, since that is the smallest number into which 4, 5 and 6 all divide. So the *LCM* and *GCD* provide upper and lower bounds respectively for the numbers in *Chord*. Even better – their ratio is invariant, since multiplying the notes in a Chord by the same (whole) number will increase *LCM* and *GCD* both by the same factor. This gives an invariant definition for the *Complexity* of the chord:

$$CY = Complexity = {}^{LCM}/_{GCD} \qquad \textbf{Equation 4}$$

The function *Complexity* above is very similar to Euler's intermediate step for *LCM* in his derivation of the sweetness function (*ESF*). Euler appeared to divide every term inside the *LCM* calculation by the *GCD*, whereas it is simpler to calculate the two independently and then divide them. In any case, both give the same results, but Equation 4 is more succinct. This *Complexity* function is an extension of Benedetti height from 2 notes to *N* notes, and represents the ratio of the largest to the smallest structures in the waveform of the chord. It is fundamental in understanding the structure of harmony.

For unison (*N*=1), *Complexity* evaluates to 1. For a 2-note *Chord*, the *Complexity* is the same as *BenedettiHeight*, so for a reduced ratio $a/b$ then $CY = a{\cdot}b$, e.g. for a perfect fifth (3:2) then $CY = 6$. For three or more numbers in *Chord*, the full formula for *CY* calculated from *LCM* and *GCD* ought to be used.

Generally speaking, more 'consonant' chords have low *Complexity* values, and more 'dissonant' chords have high *Complexity* values; the value can be arbitrarily high so there is no limit on dissonance, whereas consonance has limited number of chord arrangements. This provides a good mathematical explanation for why we prefer only a few chords (such as major or minor triads) out of all possible chords: our preferred chords have low *Complexity* value, whereas other chords have higher *Complexity* value. However, if the *Complexity* value is too low (e.g. an octave 2/1 has value 2; a perfect fourth 4/3 has value 12) then the harmony is too simple. A hypothesis would be that there is a happy middle ground with *Complexity* not too high, not too low, where the aesthetic value of harmony is maximised; a caveat being that this is probably dependent on the level of musical sophistication of the culture and observer.

Typically for chords with 3 notes, if the chord is consonant the *Complexity* will be below 1000, but if the chord is dissonant the *Complexity* will be above 1000; this cut-off point increases for more notes (larger *N*); further work could investigate how this subjective cut-off point is affected by how many notes there are.

**Table 1: Evaluation of *Complexity* function, and its base-2 logarithm *LogComplexity*, for various chords**

| Chord Description | *Chord* *CH* | *GCD* | *LCM* | *Complexity* *CY* | *LogComplexity* *LCY* |
|---|---|---|---|---|---|
| Unison | 1 | 1 | 1 | 1 | 0.0000 |
| Unison (each note ×3) | 3 | 3 | 3 | 1 | 0.0000 |
| Octave | 1, 2 | 1 | 2 | 2 | 1.0000 |
| Perfect Fifth | 2, 3 | 1 | 6 | 6 | 2.5850 |
| Perfect Fourth | 3, 4 | 1 | 12 | 12 | 3.5850 |
| Perfect Fourth (×5) | 15, 20 | 5 | 60 | 12 | 3.5850 |
| Major Third | 4, 5 | 1 | 20 | 20 | 4.3219 |
| Minor Third | 5, 6 | 1 | 30 | 30 | 4.9069 |
| Major Whole Tone | 8, 9 | 1 | 72 | 72 | 6.1699 |
| Minor Whole Tone | 9, 10 | 1 | 90 | 90 | 6.4919 |
| Major triad (root position) | 4, 5, 6 | 1 | 60 | 60 | 5.9069 |
| Major triad (first inversion) | 5, 6, 8 | 1 | 120 | 120 | 6.9069 |
| Major triad (second inversion) | 6, 8, 10 | 2 | 120 | 60 | 5.9069 |
| Major triad (all numbers ×2) | 8, 10, 12 | 2 | 120 | 60 | 5.9069 |
| Minor triad | 10, 12, 15 | 1 | 60 | 60 | 5.9069 |
| Major chord (spread-out voicing) | 1, 3, 5 | 1 | 15 | 15 | 3.9069 |
| Supermajor triad | 14, 18, 21 | 1 | 126 | 126 | 6.9773 |
| Ultramajor triad | 10, 13, 15 | 1 | 390 | 390 | 8.6073 |
| Dominant Seventh | 4, 5, 6, 7 | 1 | 420 | 420 | 8.7142 |
| Neutral triad | 18, 22, 27 | 1 | 594 | 594 | 9.2143 |
| Wolf triad | 27, 32, 40 | 1 | 4320 | 4320 | 12.0768 |
| Augmented | 12, 15, 19, 24 | 1 | 2280 | 2280 | 11.1548 |
| Diminished | 10, 12, 14, 17, 20 | 1 | 7140 | 7140 | 12.8017 |

In Table 1 *Complexity* has been evaluated for several interesting chords; for unison the value is always 1; for an octave it is 2; for major and minor triads it is 60 or 120 (depending on the inversion, and is of the form $2^n15$ depending on the voicing); for a wolf triad it is much higher (indicating dissonance); multiplying each note by a constant does not affect the *Complexity* value since it is invariant, this is demonstrated three separate times in the table above. Values for *LogComplexity* (see Equation 8) are also given, which is the *N*-note analogue for Tenney height, the base-2 logarithm of Benedetti height. Note that the chord's *Complexity* value *does not* describe the chord in full – major and minor triads have the same *Complexity*, but sound different. A chord (with *GCD* = 1) is however described in full by which factors it uses out of all the possible factors of *Complexity*, the subject of the next section.

## 6) The *ComplexitySpace* lattice of factors

*Complexity* is a positive whole number, and has a set of factors which are all the numbers (between 1 and *CY*) which divide into *CY*. This is the 'complete' chord described by Euler (1739). Denote this set of numbers *CYS* = *ComplexitySpace*. An example: if *CY* = 10, *CYS* = (1, 2, 5, 10). This set is also an invariant of *Chord*, and expands any *Chord* (with *GCD* = 1) into its complete chord.

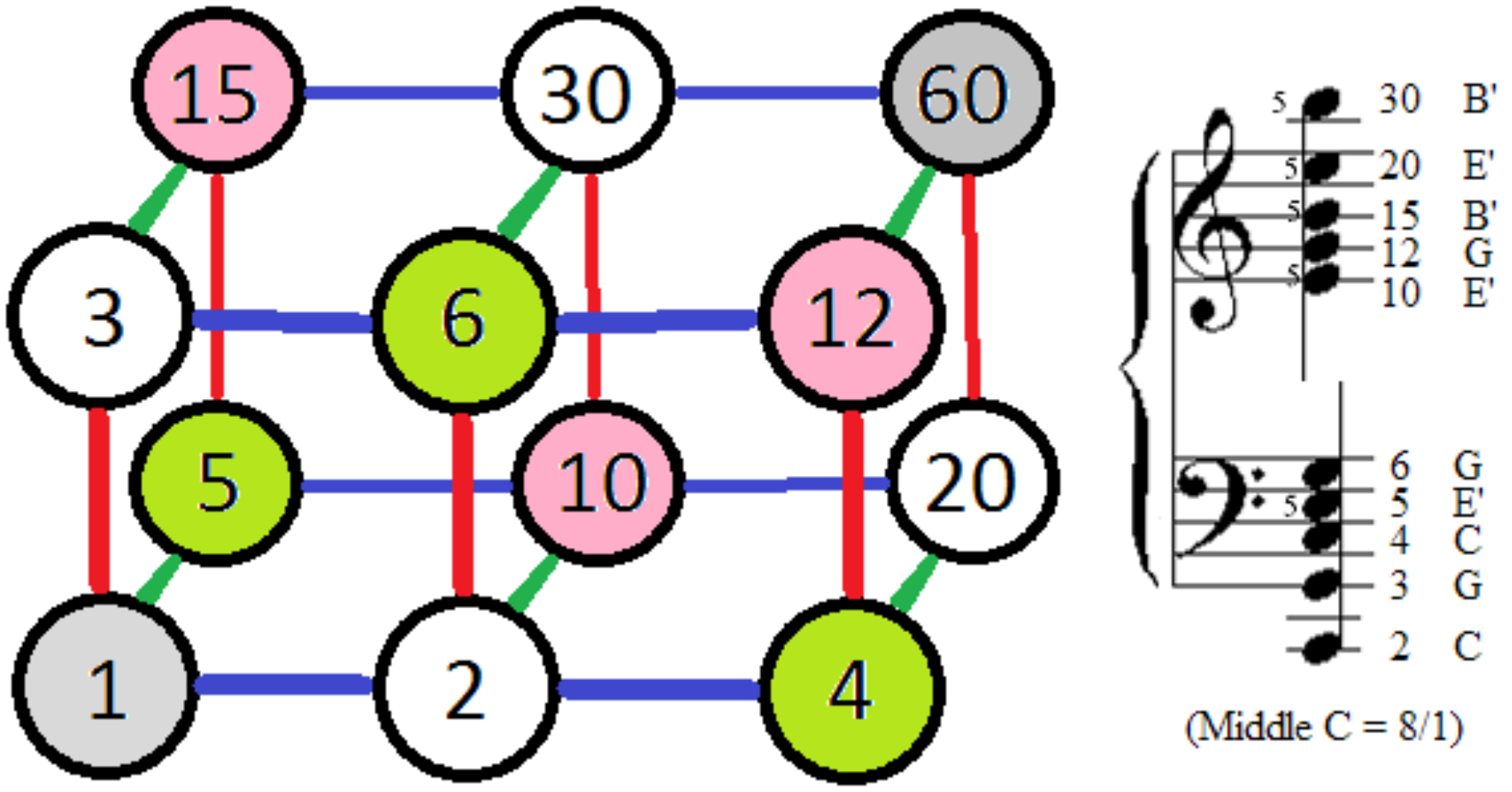

**Figure 6: a) on left – diagram of *ComplexitySpace* divisor lattice for the *Complexity* value 60, including major triad (4:5:6) in light green, and minor triad (10:12:15) in pink. Links are multiplying/dividing by 2 (blue), 3 (red) and 5 (green). b) on right – factors from 2 to 30 illustrated on a stave with pitch classes specified**

Let *p* be the highest prime which divides the *Complexity* value. If *p* is the *n*th prime then *ComplexitySpace* is a subset of the *n*-dimensional *p*-limit tone lattice. The set of divisors take up a (possibly higher-dimensional) rectangular section of the Tonnetz. The tone lattice lines give a multiplicative structure to the divisor set, where a line connects any two divisors if their ratio is a prime number. The divisor set also has a natural ordering from least (1) to greatest (*CY*). The example in Figure 6a is for *Complexity* = 60; the numbers 3 and 6 are linked since their ratio is 2 which is a prime factor of 60; all chords which have a *Complexity* of 60 will be a subset of this lattice; major (4, 5, 6) and minor (10, 12, 15) chords are highlighted. Chords can be analysed in terms of where they fall on this lattice, (see below) on measurements with respect to prime numbers.

For a given *ComplexitySpace* the notes can be plotted on a musical stave, which helps trained musicians understand the harmony. This has been done in Figure 6b using a piano stave with treble and bass clefs. To provide a reference point, a base frequency must be chosen which matches a note from the stave to a suitable frequency, and enables the elements of *ComplexitySpace* to fit on the musical stave. Matching the stave note Middle C to the frequency 8/1 gives enough space to display factors of 60 from 2 to 30 which require almost four octaves on the stave. (It doesn't matter that 8 is not a factor of 60, this just means that Middle C is not a stave note in this example.) Each numerical factor of 60 is displayed in the centre of Figure 6b with a stave note on the left and an ASCII notation on the right. The ASCII notations for pitch classes (notes up to octave equivalence), for example C = {…1/4, 1/2, 1/1, 2/1, 4/1, 8/1…} and E' = {…5/8, 5/4, 5/2, 5/1, 10/1, 20/1…}, have been described elsewhere (Ryan 2016). As Figure 6b above indicates, information regarding higher primes such as '5' can be used as accidentals on a stave tuned to 3–limit (Pythagorean) notes. This enables staves to specify free–JI compositions precisely – a topic which is expected to be discussed at greater length in future work.

**Table 2: Evaluation of position of major and minor triads with respect to their *ComplexitySpace*, the factors of 60**

| Chord Description | Divisors of 60 in their natural order (chosen subset highlighted) |
|---|---|
| All divisors of 60 | 1, 2, 3, 4, 5, 6, 10, 12, 15, 20, 30, 60 |
| Major triad | 1, 2, 3, 4, 5, 6, 10, 12, 15, 20, 30, 60 |
| Minor triad | 1, 2, 3, 4, 5, 6, 10, 12, 15, 20, 30, 60 |

In Table 2 the *ComplexitySpace* for 60 is shown in its natural order, and chords are analysed by their position in this order. The major triad is seen to be nearer 1 than 60, and vice-versa for the minor triad. Studying how to measure this in an invariant way will lead to precise measurements for how 'otonal' or 'utonal' a chord is.

## 7) *Otonality* and *Utonality* in relation to the *ComplexitySpace*

The concepts of otonality and utonality can be made precise, measured as properties of a *Chord* within its invariant *ComplexitySpace*. If a chord mainly uses the lower members of its *ComplexitySpace* (such as 4:5:6 in Table 2) then it can generally be described by smaller whole numbers (e.g. 4, 5, 6) than reciprocal numbers (e.g. 15, 12 and 10 in 60/15, 60/12, 60/10). This is the property of positive *Otonality*. Conversely, if a chord is near the top of its *ComplexitySpace* (such as 10:12:15 in Table 2) then it can generally be described by small whole reciprocal numbers (e.g. 6, 5, 4 in 60/6, 60/5, 60/4, which can have key transposed to 1/6, 1/5, 1/4) which are smaller than the original numbers (e.g. 10, 12, 15). This is the property of positive *Utonality*.

If a chord is exactly in the middle (certainly 4:5:6:10:12:15 would be) then it is neither otonal or utonal, and should measure zero on both functions. In fact, otonal and utonal are opposites (dual properties), so *OTC = Otonality = –Utonality = -UTC* (the C stands for 'coefficient'). Also both otonal and utonal measures are coefficients, so should be within the range [-1, 1] with the extremes representing a chord which is maximally otonal or utonal.

The *ComplexitySpace* has a multiplicative structure, and by taking logarithms of its elements (base-2, so octaves map to '+1') we get an additive structure, *LogComplexitySpace,* on which mean-averages can be taken to get accurate (logarithmic) values of position. For example, in *ComplexitySpace* the mean-average of 1 and 60 is 30.5, which is not a good average of the factors of 60. However, in *LogComplexitySpace* the mean-average of log(1) and log(60) is (log(1)+log(60))/2 = log(60)/2 = log($60^{1/2}$) which is approximately log(7.746); 7.746 is a good average of the divisors of 60, since half are less than it and half are greater than it. Hence logarithms base-2 will be used frequently when measuring 'averages' of chords. And the position of this average within its bounds is what will be used to measure *Otonality*.

## 8) Functions leading to definition of *Otonality* and *Utonality*

From above: *Chord* has *N* notes and describes a chord/harmony to analyse, which are positive whole numbers in increasing order. Also, averages of position will be taken on a log-scale. Here are a series of functions leading to a definition for *Otonality*:

$$LCH = LogChord = \log_2(CH) \quad \textbf{Equation 5}$$

$$LGCD = LogGCD = \log_2(GCD) \quad \textbf{Equation 6}$$

$$LLCM = LogLCM = \log_2(LCM) \quad \textbf{Equation 7}$$

$$LCY = LogComplexity = \log_2(CY) = LLCM - LGCD \quad \textbf{Equation 8}$$

$$LM = LogMidpoint = \frac{Sum\ (LCH)}{N} - LGCD \quad \textbf{Equation 9}$$

The equations above re-express *Chord*, *GCD*, *LCM*, *Complexity* functions in terms of their base-2 logarithms, and a new function *LogMidpoint* is added which is the average chord position within *LogComplexitySpace* (a *Midpoint* function could also be derived by base-2 exponentiation of *LogMidpoint*). *LogComplexity* and *LogMidpoint* are invariant; *LogGCD* and *LogLCM* are not.

It is true that 0 ≤ *LogMidpoint* ≤ *LogComplexity*, however the extreme values cannot be achieved. For example if *LogMidpoint* was zero, the sum of *LogChord* would equal *N*×*LogGCD*. However, if *GCD* < *LCM*, some *Chord* values must also exceed *GCD*, which would make the sum of *LogChord* greater than *N*×*LogGCD*, which gives a contradiction. Hence *Otonality* will be defined as how large *LogMidpoint* is with respect to *LogComplexity*, with an adjustment factor for the minimum possible value of *Midpoint* (which depends on *N*):

$$OTC = Otonality = \frac{N}{N-2}\left(\frac{LCY - 2 \cdot LM}{LCY}\right) = -Utonality = -UTC \quad \textbf{Equation 10}$$

The above definition only works for *N* ≥ 3; otherwise define both functions as zero, since 1 or 2 note chords are as otonal as they are utonal. *Utonality* is always the negation of *Otonality*, and both range between -1 and 1. It is possible to obtain a perfectly otonal chord; any set of three or more coprime numbers (e.g. 3, 4, 5; or 7, 11, 13) is otonal using this definition. From these a perfectly utonal set can be constructed by dividing the product of the whole set by each individual number in turn (e.g. 12, 15, 20; or 77, 91, 143).

*Otonality* and *Utonality* are related to the mean average of the (logarithmic) positions of the chord's notes in its *ComplexitySpace*. It is also possible to construct higher order functions such as standard deviation and skewness:

$$SPC = SpreadCoeff = \sqrt{\frac{4}{N}\sum_{k=1}^{N}\left(\frac{LCH(k) - LM - LGCD}{LCY}\right)^2} \quad \textbf{Equation 11}$$

Here the *SpreadCoeff* is an invariant function in the range 0 to 1 which describes how spread out the notes are in *ComplexitySpace*: a value of 1 can be achieved by the chord (1, 2) which is maximally spread out from its midpoint (on a log-scale) which is $2^{1/2}$; a *SpreadCoeff* value near to 0 can be achieved by a chord ($k$, $k$+1) for large $k$. since the frequency ratio is small compared to the *Complexity* value of $k(k+1)$. The chord (1, $k$) has *SpreadCoeff* value of 1 (for $k$>1) however for $N$>3 the value 1 is not achievable – further work would be to find a function of $N$ to include in *SpreadCoeff* which allows the maximum of 1 to be attained for all $N$.

$$SK = Skewness = \sqrt[3]{\frac{1}{N}\sum_{k=1}^{N}\left(\frac{LCH(k) - LM - LGCD}{LCY}\right)^3} \qquad \textbf{Equation 12}$$

Higher order functions such as *Skewness* above, or kurtosis etc, can be constructed from sums of higher powers of the bracketed expression. Further work would be to construct these functions with suitable normalisation constants (dependent on both the order and on $N$) which give coefficients across the whole range [-1, 1] for odd orders and [0, 1] for even orders. *Skewness* as defined above has been evaluated to a range of only [0.000, 0.458] so could be improved.

**Table 3: Some examples of the functions in this section. Black rows are invariant functions, grey rows are non-invariant functions.**

| ***CH*** | ***Chord*** | 4, 5, 6 | 12, 15, 18 | 5, 7, 11, 13, 21 | 3, 4, 5 | 12, 15, 20 | 1, 2, 30, 60 | 1, 30, 60 |
|---|---|---|---|---|---|---|---|---|
| ***N*** | ***N*** | 3 | 3 | 5 | 3 | 3 | 4 | 3 |
| ***GCD*** | ***GCD*** | 1 | 3 | 1 | 1 | 1 | 1 | 1 |
| ***LCM*** | ***LCM*** | 60 | 180 | 15015 | 60 | 60 | 60 | 60 |
| ***CY*** | ***Complexity*** | 60 | 60 | 15015 | 60 | 60 | 60 | 60 |
| ***LCY*** | ***LogComplexity*** | 5.9069 | 5.9069 | 13.8741 | 5.9069 | 5.9069 | 5.9069 | 5.9069 |
| ***LM*** | ***LogMidpoint*** | 2.3023 | 2.3023 | 3.3363 | 1.9690 | 3.9379 | 2.9534 | 3.6046 |
| ***OTC*** | ***Otonality*** (coefficient) | 0.6614 | 0.6614 | 0.8651 | 1.0000 | -1.0000 | 0.0000 | -0.6614 |
| ***SPC*** | ***SpreadCoeff*** | 0.0810 | 0.0810 | 0.1034 | 0.1021 | 0.1021 | 0.8478 | 0.8740 |
| ***SK*** | ***Skewness*** | -0.0201 | -0.0201 | 0.0139 | -0.0273 | 0.0273 | 0.0000 | -0.3743 |

The functions developed up to this point have been calculated in Table 3 for a range of chords. The chords have been chosen to give as wide a variety in values of the functions as possible, e.g. for a coefficient to demonstrate (where possible) both minimum and maximum values. A complete function list from this paper can be found in Appendix 1, with some more example calculations in Appendix 2 where dominant seventh chords and various ninth chords are compared, with example scores given.

## 9) Invariant functions on ratios between (ordered) *Chord* values

The values of *Complexity* and *Otonality*, and other functions derived above, are properties of the entire chord, global properties. Other types of invariant function are local properties, between two notes at a time, or determined by a limited number of the notes such as maximum and minimum values. To derive these, the *Chord* will be ordered by ascending frequency. Each individual note is not invariant, but the frequency ratio between any two different notes **is** invariant. This section presents invariant functions based on ratios of the notes in the *Chord*.

$$CR(m,n) = Ratio(m,n) = \frac{CH(n)}{CH(m)} \quad \text{for } 1 \le m,n \le N$$ **Equation 13**

$$CR(k) = Ratio(k) = CR\ (k, k+1) = \frac{CH(k+1)}{CH(k)} \quad \text{for } 1 \le k < N$$ **Equation 14**

*Ratio*($k$) measures the frequency ratio between consecutive notes in the chord, and the set of these determines the whole chord. The *Ratio* functions above are all invariant, and with *Chord* in ascending order the *Ratio*($k$) values will be greater than 1. (Equality to 1 would only be possible if *Chord* repeated a note, however the assumption so far is that the notes of Chord are distinct.)

$$MNR = MinRatio = \min(\ CR(k) : 1 \le k < N\ )$$ **Equation 15**

$$MXR = MaxRatio = \max(\ CR(k) : 1 \le k < N\ )$$ **Equation 16**

$$TR = TotalRatio = \frac{CH(N)}{CH(1)}$$ **Equation 17**

Three more invariant functions: *MinRatio* is the smallest ratio of consecutive frequencies in the chord; *MaxRatio* is the largest ratio of consecutive frequencies in the chord; *TotalRatio* is the frequency ratio between the first and last (lowest and highest) notes in the chord.

As before, taking logarithms base-2 is a good idea; it preserves invariance, it turns a multiplicative structure into an additive structure, and turns an octave difference into an 'add 1'. Here are the same functions in logarithmic form:

$$LCH(n) = LogChord(n) = \log_2(\ CH(n)\ ) \quad \text{for } 1 \le n \le N$$ **Equation 18**

$$LCR(m,n) = LogRatio(m,n) = LCH(n) - LCH(m) \quad \text{for } 1 \le m,n \le N$$ **Equation 19**

$$LCR(k) = LogRatio(k) = LCR(k,k+1) = LCH(k+1) - LCH(k) \quad \text{for } 1 \le k < N$$ **Equation 20**

$$LMNR = LogMinRatio = \log_2(MNR) = \min(\ LCR(k) : 1 \le k < N\ )$$ **Equation 21**

$$LMXR = LogMaxRatio = \log_2(MXR) = \max(\ LCR(k) : 1 \le k < N\ )$$ **Equation 22**

$$LTR = LogTotalRatio = \log_2(TR) = LCH(N) - LCH(1)$$ **Equation 23**

Ranges for the final three: since there are ($N$-1) ratios, *LogMinRatio* must be between 0 and *LogTotalRatio*/($N$-1); *LogMaxRatio* must be between *LogTotalRatio*/($N$-1) and *LogTotalRatio*; *LogTotalRatio* must be between 0 and *LogComplexity*. These ranges can be used to produce coefficients in the range [0, 1] for max, min and total ratios:

$$MNRC = MinRatioCoeff = (N-1)\frac{LMNRo}{LTR}$$ **Equation 24**

$$MXRC = MaxRatioCoeff = \frac{N-1}{N-2}\left(\frac{LMXR}{LTR} - \frac{1}{N-1}\right)$$ **Equation 25**

$$TRC = TotalRatioCoeff = \frac{LTR}{LCY}$$ **Equation 26**

These three coefficients measure the size of the min, max or total ratio compared to its minimum and maximum possible sizes. Examples are given in the table below:

**Table 4: Examples of maximum and minimum values for three invariant chord coefficients**

| Function | Chords with minimum value 0 | Interpretation of minimum value | Chords with maximum value 1 | Interpretation of maximum value |
|---|---|---|---|---|
| *MinRatioCoeff* *MNRC* | ($n$, $n$+1, 2$n$) as $n\to\infty$ | Minimum frequency ratio tends to zero compared to whole chord | 1, 2, 4<br>1, 2, 4, 8<br>1, 2, 4, 8, 16<br>5, 15, 45<br>etc. | All ratios equal minimum ratio, so pitches of notes equally spread<br>(Pitch is log-frequency) |
| *MaxRatioCoeff* *MXRC* | 1, 2, 4<br>1, 2, 4, 8<br>3, 6, 12<br>etc. | All ratios equal maximum ratio, so pitches equally spread | ($n$, $n$+1, 2$n$) as $n\to\infty$ | Maximum ratio tends to the same size as the whole chord |
| *TotalRatioCoeff* *TRC* | ($n$, $n$+1) as $n\to\infty$ | Chord uses vanishing amount of full range 1 to Complexity which is [1, n(n+1)] | (1, $n$) as $n\to\infty$ | Chord fills its full range 1 to Complexity which is [1, $n$] |

The ratios are interrelated – if the MinRatioCoeff is large, then the MaxRatioCoeff tends to be smaller. Together these functions give useful information about how the chord is spread out.

## 10) Invariant functions with respect to particular prime numbers

Many of the functions above will also work when projecting the original chord onto the space spanned by one or more prime numbers. A prime number $p$ is a number with exactly two factors; 1 and $p$; the sequence of primes starts 2, 3, 5, 7, 11, 13, 17, 19, 23…. The space spanned by each prime is its set of powers $p^n$ for $n$ a zero or positive whole number; for example the space spanned by 2 is 1, 2, 4, 8, 16…, and the space spanned by 13 is 1, 13, 169, 2197…. The space spanned by multiple primes are the numbers obtained by multiplying together any positive (or zero) powers of each prime; for example the space spanned by 5 and 7 starts 1, 5, 7, 25, 35, 49, 125…

Given a set of prime numbers *P* to project onto, and a chord to project (e.g. major triad 4:5:6): first factorise the chord (4:5:6 = 2×2 : 5 : 2×3) and the replace each prime factor *p* by either 1 (if prime *p* is not in the set *P*) or leave *p* alone (if *p* is in *P*). For example, suppose *P* = {2, 3}; then 4:5:6 projects onto 2×2 : 1 : 2×3 = 4:1:6. Some more examples are given in the table below.

**Table 5: Prime projections of dominant seventh chord 4:5:6:7 = 2×2:5:2×3:7**

| Set of primes, *P* | Projected ratio (factorised) | Projected Ratio | $GCD_P$ | $LCM_P$ | $Complexity_P$ $CY_P$ |
|---|---|---|---|---|---|
| All primes | 2×2 : 5 : 2×3 : 7 | 4:5:6:7 | 1 | 420 | 420 |
| 2 | 2×2 : 1 : 2×1 : 1 | 4:1:2:1 | 1 | 4 | 4 |
| 3 | 1×1 : 1 : 1×3 : 1 | 1:1:3:1 | 1 | 3 | 3 |
| 5 | 1×1 : 5 : 1×1 : 1 | 1:5:1:1 | 1 | 5 | 5 |
| 7 | 1×1 : 1 : 1×1 : 7 | 1:1:1:7 | 1 | 7 | 7 |
| 11 | 1×1 : 1 : 1×1 : 1 | 1:1:1:1 | 1 | 1 | 1 |
| 2, 5 | 2×2 : 5 : 2×1 : 1 | 4:5:2:1 | 1 | 20 | 20 |
| Odd primes (all except 2) | 1×1 : 5 : 1×3 : 7 | 1:5:3:7 | 1 | 105 | 105 |
| Primes higher than 3 | 1×1 : 5 : 1×1 : 7 | 1:5:1:7 | 1 | 35 | 35 |

In Table 5 some examples of projection are given for a dominant seventh chord 4:5:6:7, which uses only the primes 2, 3, 5, 7. Since *P* is the subset (of all the primes) to project onto, a new invariant function $Complexity_P$ is obtained. We can check that the previous *Complexity* value (420) is the same as the product of $Complexity_P$ across primes 2, 3, 5, 7 (indeed it is, since 4×3×5×7 = 420). In general, *Complexity* can be decomposed geometrically in terms of $Complexity_p$ for each individual prime direction *p* involved in the chord. This is related to *ComplexitySpace* being rectangular on the Tonnetz; its size in each *p*-direction is independent of all the other directions and equal to $CY_p$.

If *P* is the set of all prime numbers then $Complexity_P$ = *Complexity*. Indeed, if *P* is the set of all prime numbers then projecting any invariant function onto it gives the same function back. More interesting cases have *P* as a proper subset of all the primes.

It could be attempted to derive prime projections of other functions. However care ought to be taken, for after prime-projection *Chord* may no longer have distinct values or retain its order. *Otonality* assumed that all *Chord* notes are distinct, but this is no longer always true after projection. Functions like *Midpoint* and *Otonality* could only be prime-projected after repeated notes are dealt with correctly, for which some work is presented below in the section on weighted invariant functions.

Some functions only make sense when the original chord is ordered. The ratio-based functions defined above (section 9) depend on *Chord* being listed in ascending order as *Chord*(*k*) for *k*=1..*N*. Hence when prime projections are taken, $Chord_P$ is no longer in ascending order, and $Ratio_P(m,n)$ no longer makes sense. So prime projections of the *Ratio*-based functions are not given here.

Near the end of Table 5, $Complexity_P$ is calculated for *P* the set of all odd prime numbers, e.g. the set of prime numbers excluding 2. This is a special function (denote it *OddComplexity*, abbreviated *OCY*) since by discarding all powers of 2 we are effectively considering the chord up to octave equivalence, i.e. the pitch classes within the chord. (See also the Kees height, however *OCY* has the benefit of being prime limit based, instead of odd limit based.) A function $Complexity_2$ or $CY_2$ also exists, which is simply the power of 2 in *CY*, and is equal to *CY*/*OCY*. These (*OCY*, $CY_2$) are important in the analysis of scales which repeat every octave, and will be discussed below. More generally, $CY_p$ for a specific prime number *p* is the highest prime power $p^k$ which divides into *CY*.

**Table 6: Analysis of a particular Bohlen-Pierce scale over a tritave (1/1 to 3/1)**

| | |
|---|---|
| **Scale (fractions)** | 1/1, 35/27, 7/5, 5/3, 9/5, 7/3, 25/9, 3/1 |
| **Scale (integers)** | 135, 175, 189, 225, 243, 315, 375, 405 |
| **Octave = 1200**<br>**Pitch in cents**<br>**1200 × $\log_2(f)$** | 0, 449, 583, 884, 1018, 1467, 1769, 1902 |
| **Tritave = 1300**<br>**Bohlen-Pierce pitch**<br>**1300 × $\log_3(f)$** | 0, 307, 398, 604, 696, 1003, 1209, 1300 |
| ***Complexity***<br>***CY*** | 212 625 = $3^5 \cdot 5^3 \cdot 7 \cdot$ |
| ***OddComplexity***<br>***OCY*** | 212 625 = $3^5 \cdot 5^3 \cdot 7 \cdot$ |
| ***BohlenPierceComplexity***<br>***BPCY*** | 875 = $5^3 \cdot 7 \cdot$ |
| $CY_2$, $CY_3$ | 1, $3^5$ |

Not all scales are based on the interval of an octave (2/1); Bohlen-Pierce scales are based on the interval of a tritave (3/1) and only uses odd-numbered frequencies, most commonly some combinations of the primes 3, 5 and 7. It is possible to define a *BohlenPierceComplexity* value which is $Complexity_P$ for *P* the set of primes excluding 2 and 3. In Table 6 a B-P scale is specified, and since all the whole numbers are odd, *CY* = *OCY* = 212 625, and by discarding powers of 3 a BPCY value of 875 is obtained. *BPCY* is thus a complexity value **up to tritave equivalence** for chords with only odd-numbered frequencies. *BPCY* also exists for chords with even-numbered frequencies, and corresponds to *CY* with the information about primes 2 and 3 discarded. $CY_3$ is then the ratio *OCY*/*BPCY*.

*OCY* and *BPCY* seem to be the most important functions of the form $Complexity_P$ where *P* is a proper subset of all prime numbers. Out of these two, *OCY* seems to be slightly more useful since in humans the ear hears pitch classes (octave-equivalent) as the same note, but tritaves sound like different notes. Nonetheless, *BPCY* is a useful statistic regarding the non-Pythagorean content of a harmony.

## 11) Dealing with multiplicity of notes, and amplitude-weighted chords

Initially it was assumed that all note frequencies were distinct. However, there are various motivations for considering repeated notes. Firstly, when multiple instruments play music, the same note may be repeated by more than one instrument. Secondly, distinct notes can become repeated under prime projections. Thirdly, the theory to this point has not taken into account the loudness or volume of each frequency, and doubling the loudness has the same effect on the waveform as playing the note on two channels (albeit perfectly in phase). Hence it is necessary to consider the general case of repeated frequencies.

When a note occurs more than once, this is the same as listing the distinct notes and giving them whole-numbered multiplicities. E.g. the frequency set {10, 10, 10, 12, 15} can be represented by the distinct frequencies (10, 12, 15) with the multiplicities (3, 1, 1). Since when a note is played two or more times simultaneously (in phase) it becomes louder, if the multiplicities are allowed to be any positive number then they specify how loud each note is. This set of multiplicities can therefore be referred to in various ways: as amplitudes, loudnesses, volumes or weightings.

The frequencies in the chord have already been described as $CH = Chord = \{CH(i) : i=1..N\}$, so now define the weightings $WT = Weight = \{WT(i) : i=1..N\}$. When weights are introduced, some functions are unchanged, but others are changed. If a function changes, it will be prefixed by *Weighted*- or *W*- to show the alternative definition. For such a function, it is desirable for it to be continuous in all $WT(i)$, i.e. when a weighting changes by a small amount, the function also changes by only a small amount. Weights are now potentially any positive real number instead of just whole numbers. Caution ought to be taken if any weight goes to zero, since some functions may be discontinuous at zero weight; for example, *Complexity* may have a step change if a note goes to zero weight and if it is considered to be removed from the chord.

As the parameters of *Weight* vary, some functions remain unchanged. These include: *Chord*, *N*, all *Ratio* derived functions, *GCD*, *LCM*, *Complexity*, *ComplexitySpace*, and all their logarithms. They are invariant since they operate on the (distinct) *Chord* values and not on the *Weight* values. It is assumed that zero volume notes count towards the *Complexity* calculation.

Here are some (re-)definitions of functions which change when weightings are introduced:

$$SWT = SumWeight = \sum_{i=1}^{N} WT(i) \qquad \textbf{Equation 27}$$

$$WLM = WeightedLogMidpoint \ = \frac{\sum[WT(i)\ LCH(i)]}{SWT} - \ LGCD \qquad \textbf{Equation 28}$$

$$WOTC = WeightedOtonality = \left(\frac{LCX - 2WLM}{LCX}\right) = -WeightedUtonality = -WUTC \qquad \textbf{Equation 29}$$

When a *Chord* has an associated *Weight*, invariant properties should be unchanged if all the *Weight* elements are multiplied by a constant factor, i.e. independent of how loud or quiet the *Chord* is played. By this criteria, *Weight* itself is not invariant, but ratios between its elements are invariant. *SumWeight* is also not invariant, however *WeightedLogMidpoint* and *WeightedOtonality* are invariant. *WeightedLogMidpoint* appears to be the basic function invariant of both *Weight* and *Chord* (i.e. multiplying either by a constant factor), which is achieved by first taking a weighted average (which cancels out constant factors in *Weight*) and then by subtracting *LogGCD* (which cancels out constant factors in *Chord*).

In Equation 27 and Equation 28 above, all sums are across the *N* distinct notes. *SumWeight* would be the largest possible amplitude of the note combination played by sine waves with frequencies specified by *Chord* and amplitudes specified by *Weight*. The new coefficient for *WeightedOtonality* drops the factor *N*/(*N*-2) which was present in the unweighted *Otonality* function, since the extreme values have changed and the coefficient needs to stay in the range [-1, 1]. (A brief explanation would be that a chord of the form (1, 4, 5, 6) could now have a weighting of the form (*w*, 1, 1, 1) putting most of the weight *w* on the frequency 1, so the *WeightedLogMidpoint* has a larger range (for fixed *N*) than the *LogMidpoint*.) In addition, higher functions (*WeightedSpreadCoeff*, *WeightedSkewness*, etc.) could also be derived if needed.

As for prime-projections of weighted chords, such a definition would only be useful if the function both behaved well under prime projections, and varied with the weighting function. Since none of *Complexity*, *Ratio* and *Otonality* functions meet both conditions, this subject is not considered further here.

**Table 7: Some examples of weighted function calculations (black for invariant, grey for non-invariant)**

| ***Chord*** ***CH*** | (4, 5, 6) | (4, 5, 6) | (4, 5, 6) | (2, 3, 4, 5, 6, 7) | (2, 3, 4, 5, 6, 7) |
|---|---|---|---|---|---|
| ***Weight*** ***WT*** | (1, 1, 1) | (10, 1, 1) | (1, 1, 10) | (1, 1, 1, 1, 1, 1) | (10, 10, 10, 1, 1, 1) |
| ***N*** | 3 | 3 | 3 | 6 | 6 |
| ***SumWeight*** ***SWT*** | 3 | 12 | 12 | 6 | 33 |
| ***GCD*** | 1 | 1 | 1 | 1 | 1 |
| ***LCM*** | 60 | 60 | 60 | 420 | 420 |
| ***Complexity CY*** | 60 | 60 | 60 | 420 | 420 |
| ***LogComplexity LCY*** | 5.907 | 5.907 | 5.907 | 8.714 | 8.714 |
| ***LogMidpoint, LM*** (ignore weightings) | 2.302 | 2.302 | 2.302 | 2.050 | 2.050 |
| ***WeightedLogMidpoint WLM*** | 2.302 | 2.076 | 2.514 | 2.050 | 1.623 |
| ***Otonality, OTC*** (ignore weightings) | 0.661 | 0.661 | 0.661 | 0.794 | 0.794 |
| ***WeightedOtonality WOTC*** (drops $N/(N\text{-}2)$ factor) | 0.220 | 0.297 | 0.149 | 0.530 | 0.627 |

In Table 7 above some of these weighted functions have been calculated for examples based on a major triad chord (4:5:6) and an extended dominant seventh chord (2:3:4:5:6:7). *WeightedLogMidpoint* increases if the higher frequency notes of *Chord* receive higher weightings, and decreases if the lower frequency notes have more weight. *WeightedOtonality* moves in the opposite direction to *WeightedLogMidpoint* under these changes, since more weight on the lower notes makes a note combination more otonal. Note that *WeightedOtonality* depends to a large extent on what the *Complexity* is considered to be: for a combination of *Chord* = (4, 5, 6) and *Weight* = (1, 1, 0), the *Complexity* would be 60 using the scheme above, but 20 if the chord was converted back into an equivalent unweighted form. The *WeightedOtonality* calculation gives different values for these different *Complexity* values. The proposed solution to this issue would be to compare *WeightedOtonality* only if the *Complexity* values are the same. This makes *WeightedOtonality* more limited in scope than *WeightedLogMidpoint*, which is comparable even if *Complexity* values are different.

## 12) Analysis of higher harmonics of notes in a chord

In Table 7 the chord 2:3:4:5:6:7 was used in calculations with two different weightings. This suggests an application for measurements on weighted chords: to analyse harmonic series for different non-sinusoidal waveforms. Each waveform has a Fourier expansion in terms of waves of whole-numbered frequency, where the (complex) amplitude of each wave determines the original waveform. By analysing a fixed set of frequencies of these Fourier expansions, *Complexity* is kept constant so *WeightedOtonality* can be compared. Note that *Weight* should be the positive real-valued magnitude of the (possibly negative or complex-valued) coefficient in the Fourier series.

**Table 8: Comparison of four simple waveforms for unit frequency over the first eight harmonics**

| **Waveform** | Sine | Triangle | Square | Sawtooth |
|---|---|---|---|---|

| **Start of Fourier Expansion** | sin(t) | sin(t) - sin(3t)/9 + sin(5t)/25… | sin(t) + sin(3t)/3 + sin(5t)/5… | sin(t) – sin(2t)/2 + sin(3t)/3… |
|---|---|---|---|---|
| ***Chord*** ***CH*** | (1, 2, 3, 4, 5, 6, 7, 8) | (1, 2, 3, 4, 5, 6, 7, 8) | (1, 2, 3, 4, 5, 6, 7, 8) | (1, 2, 3, 4, 5, 6, 7, 8) |
| ***Weight*** ***WT*** | (1, 0, 0, 0, 0, 0, 0, 0) | (1, 0, 1/9, 0, 1/25, 0, 1/49, 0) | (1, 0, 1/3, 0, 1/5, 0, 1/7, 0) | (1, 1/2, 1/3, 1/4, 1/5, 1/6, 1/7, 1/8) |
| ***N*** | 8 | 8 | 8 | 8 |
| ***SumWeight*** ***SWT*** | 1.000 | 1.172 | 1.676 | 2.718 |
| ***Complexity*** ***CY*** | 840 | 840 | 840 | 840 |
| ***LogComplexity*** ***LCY*** | 9.714 | 9.714 | 9.714 | 9.714 |
| ***LogMidpoint, LM*** (ignore weightings) | 1.912 | 1.912 | 1.912 | 1.912 |
| ***WeightedLogMidpoint*** ***WLM*** | 0.000 | 0.279 | 0.832 | 1.177 |
| ***Otonality, OTC*** (ignore weightings) | 0.808 | 0.808 | 0.808 | 0.808 |
| ***WeightedOtonality*** ***WOTC*** (drops *N*/(*N*-2) factor) | 1.000 | 0.943 | 0.829 | 0.758 |

In Table 8 the *WeightedOtonality* coefficient was found to decrease across the four waveforms Sine, Triangle, Square, Sawtooth when considering the first eight coefficients. This means that these waveforms, listed in this order, had a lower proportion of lower harmonics and a higher proportion of higher harmonics present.

Calculations were also done for a sawtooth wave over the first 8, 16 and 32 harmonics; *WeightedOtonality* was found to take the values 0.758, 0.831, 0.910 respectively. As the number of harmonics goes to infinity, the limit of this series is thought to be 1.000 and hence not a particularly helpful measurement. This happens since *Complexity* increases much faster than *WeighedLogMidpoint*. Further work is needed to ascertain if *WeightedLogMidpoint* itself converges to a finite value in the infinite limit, for common waveforms such as Triangle ($1/n^2$ weighted overtones) or Sawtooth ($1/n$ overtones). Further work is also needed to analyse sounds made with multiple notes and compound waveforms. For example, what invariant functions exists for a major triad (4:5:6) played with trapezium waves? Can the interval 3:2 played with triangle waves be shown to be more consonant than 40:27 played with triangle waves? Helmholtz (1885) and Plomp & Levelt (1965) hypothesised that compound tones sound more consonant when the overtones (partials) overlap – can this be demonstrated with some kind of invariant function?

## 13) Scale analysis up to octave equivalence

After a diversion into weightings of chords, back to unweighted chords. This time consider an octave-based scale with $N$ pitch classes, where the notes (pitch class representatives) increase in frequency over the octave from a lowest frequency $k_1$ through $k_2$, $k_3$…, $k_N$ and then to a highest frequency $2k_1$. In total there are ($N$+1) notes in the complete scale, since one pitch class appears at both the bottom and top of the scale. Mathematical terminology for such a scale is a sequence ($k_1$, $k_2$, … , $k_N$, $k_{N+1} = 2k_1$) where $k_i < k_j$ if $i < j$.

For a scale over an octave, a nice property to have is for a function to not be affected by a reordering of the scale, which is moving the scale up or down. Moving the scale up means discarding $k_1$ and adding $2k_2$ to the end. Moving the scale down means discarding $2k_1$ and inserting $k_N/2$ at the start. These reorderings cycle the pitch classes around, and there are $N$ distinct reorderings. As for functions invariant with respect to reorderings, *Complexity* does not quite have this property, but *OddComplexity* does, since each reordering cycles the odd parts of the prime factorisations of the notes, without changing any of them.

**Table 9: Reorderings of the short scale (3, 4, 5, 6)**

| ***Chord* (*CH*)** | (3, 4, 5, 6) | (4, 5, 6, 8) | (5, 6, 8, 10) | (6, 8, 10, 12) |
|---|---|---|---|---|
| ***N*** | 4 | 4 | 4 | 4 |
| ***GCD*** | 1 | 1 | 1 | 2 |
| ***LCM*** | 60 | 120 | 120 | 120 |
| ***Complexity* (*CY*)** | 60 | 120 | 120 | 60 |
| ***OddComplexity* (*OCY*)** | 15 | 15 | 15 | 15 |
| ***BohlenPierceComplexity* (*BPCY*)** | 5 | 5 | 5 | 5 |

In Table 9 a complete set of reorderings are given for the (short) scale 3:4:5:6, which are repeated reorderings until the original scale is obtained again (after dividing out by the *GCD*). As a scale is reordered both *GCD* and *LCM* may change by a factor of 1 or 2. Thus *Complexity* itself, as their ratio, can stay the same, increase by a factor of 2, or decrease by a factor of 2. So *Complexity* is not quite constant for any given scale (up to cyclic reordering). A hypothesis is that the minimum and maximum Complexity values across all reorderings of any scale are either the same or vary by a factor of 2. Denote the minimum *Complexity* value across these reorderings by *mCY*, so in the example above *CY* = 60 or 120 and *mCY*=60. For scales, generally *mCY* will be used to compare relative consonance.

For measuring scales it is recommended that both *Complexity* and *OddComplexity* are calculated; that is, calculate all values of *CY*, and then derive *mCY* and *OCY*. What follows is an example showing why both are needed. A general method exists (given in a later section) to construct a scale of $N$+1 notes from any set of $N$ odd numbers. Then, for two different sets of odd numbers (1, 15) and (3, 5) the *OCY* values will both be 15, but the *mCY* value of (3, 5) is much lower. To show this, for (1, 15) the scales obtained are (8, 15, 16) and (15, 16, 30); for (3, 5) the scales are (3, 5, 6) and (5, 6, 10); *mCY* = 240 in the former case, and *mCY* = 30 in the latter case. Hence (1, 15) gives *mCY* values 8 times as large as (3, 5) does. So not all subsets of an *OddComplexitySpace* are equally valuable for creating scales, and *mCY* can be used to tell them apart.

In the following sections two methods will be given for scale construction: from sets of odd numbers, and from splitting intervals. The focus of either method should be on finding scales with the lowest possible *Complexity* and/or *OddComplexity* values; scales with optimised consonance.

## 14) Odd numbers and divisor lattice scales

The method for creating a scale from any set of *N* odd numbers is: 1) find the maximum odd number in the set; 2) multiply all the other odd numbers by a power of 2 until they are in the octave below this maximum number; 3) find the minimum value from the previous step; 4) add two times this minimum value onto the top of the scale, to give a final octave-based scale of *N*+1 frequencies. These can then be reordered, if wished.

Every odd number has a set of divisors (its *OddComplexitySpace*). For that set of divisors considered as a harmony, the *CY* and *OCY* values are both the original odd number. When that set of odd divisors is turned into a scale using the method outlined above, the *OCY* value remains the same, but the *CY* value will increase by a power of 2, and out of the reorderings an *mCY* value can be obtained. Hence the divisor set of every odd number gives a scale, at least one *CY* value, and a unique *mCY* value. Here are some examples, where all the *CY* values are stated:

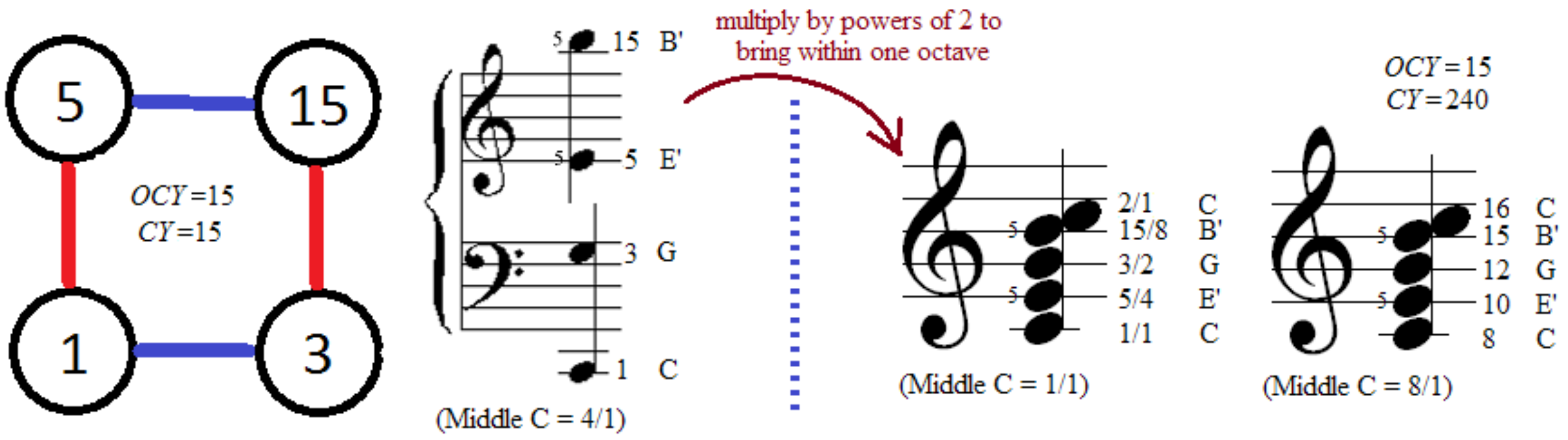

**Figure 7: a) Diagram of *OddComplexitySpace* lattice for the number 15, Links are multiplying/dividing by 3 (blue) and 5 (red). b) Factors 1, 3, 5, 15; c) as fractions between 1/1 and 2/1; d) as notes between 8 and 16**

This procedure has been carried out for the number 15 in Figure 7. Its divisor set (1, 3, 5, 15) has *CY* = *OCY* = 15, represented as a lattice (a square) in Figure 7a. Next in Figure 7b the four divisors are represented as notes on a piano stave, with the number and the name of the pitch class next to them. In Figure 7c the numbers are divided by powers of 2 to fall into the range [1, 2], giving an octave-based scale (1/1, 5/4, 3/2, 15/8, 2/1) where the scale has been completed by doubling the C from the bottom to the top. Alternatively, in Figure 7d the numbers are multiplied by powers of 2 to make them less than an octave from 15, the highest factor, giving a scale of (8, 10, 12, 15, 16). Either fractions or whole numbers can represent a scale on a stave, however the whole numbers make it easy to find *Complexity* and *OddComplexity* values. For this scale, *OCY* = 15 and *CY* = 240. Note that on the stave the base frequency (Middle C = …) can be chosen freely, and in Figure 7 above has been chosen separately in each diagram so that the numbers correspond to notes which fit on each stave.

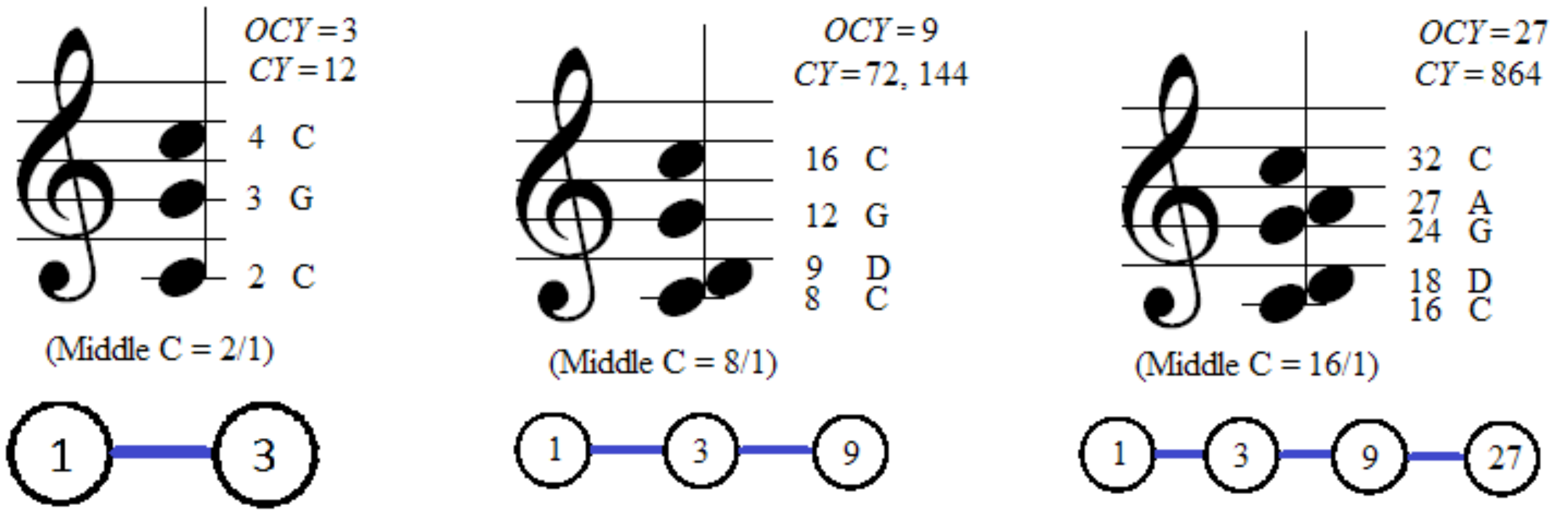

**Figure 8: Diagram of *OddComplexitySpace* lattice for the numbers a) 3; b) 9; c) 27 with associated scales**

Some simple cases for *OddComplexity* (*OCY*) values are when *OCY* has only one prime factor. In Figure 8 examples are given for 3, $9 = 3^2$ and $27 = 3^3$. Each extra power of 3 adds one extra note into the scale. *Complexity* (*CY*) value or values obtained from all reorderings are given in each diagram. Across these three diagrams, *CY* increases rapidly meaning dissonance increases rapidly too; for the last scale *CY* = 864 which is caused by the note A (27) clashing with the C (32). These scales are called 'Pythagorean' since they are generated by only the first two primes, 2 and 3. In other words, parts of Pythagorean scales start to clash when more than 3 or 4 notes are present. To solve this, higher primes are used for longer scales, typically 5 for a good major third 5/4 as in Figure 7 for *OCY* = 15.

Other scales generated by a single prime (e.g. 1, 5, 25 –> 16, 20, 25, 32 with *OCY* = 25, *CY* = 800, an augmented chord scale) could be formed, however the *CY* values for primes 5 and above are higher than the corresponding *CY* values with the prime 3. This means less consonant harmony, e.g. a three-note augmented chord is more dissonant (higher *CY*) than a three-note cycle of Pythagorean perfect fifths.

In order to minimise *CY* for small scales, two other solutions can be tried: firstly to use multiple odd primes in *OCY*, secondly to make the prime powers non-increasing in *OCY*. As examples, to use powers of both 3 and 5 instead of just 3; also to use a higher (or equal) power of 3 than 5, and likewise with powers of 5 and 7. Higher primes will generally give larger *CY* values. This explains why traditional scales focus on the prime 3, 5 and occasionally 7; to minimise *CY* and maximise scale consonance.

The second condition (non-increasing prime exponents) may not be absolutely necessary to minimise *CY*, since divisor spacing is uneven and a slightly higher prime might need a smaller power of 2 to make into a scale, and give a smaller *CY* value. For smaller scales, e.g. up to 12 notes, there are some interesting examples in which the powers are not strictly decreasing, e.g. $OCY = 3^2 \cdot 5^0 \cdot 7^1 = 63$ which gives a six-note scale. For the examples below, the rule will be that *OCY* is an odd composite number with a factor of 3 and a factor from {5, 7, 11}. The first few values meeting these conditions are (15, 21, 33, 45, 63…).

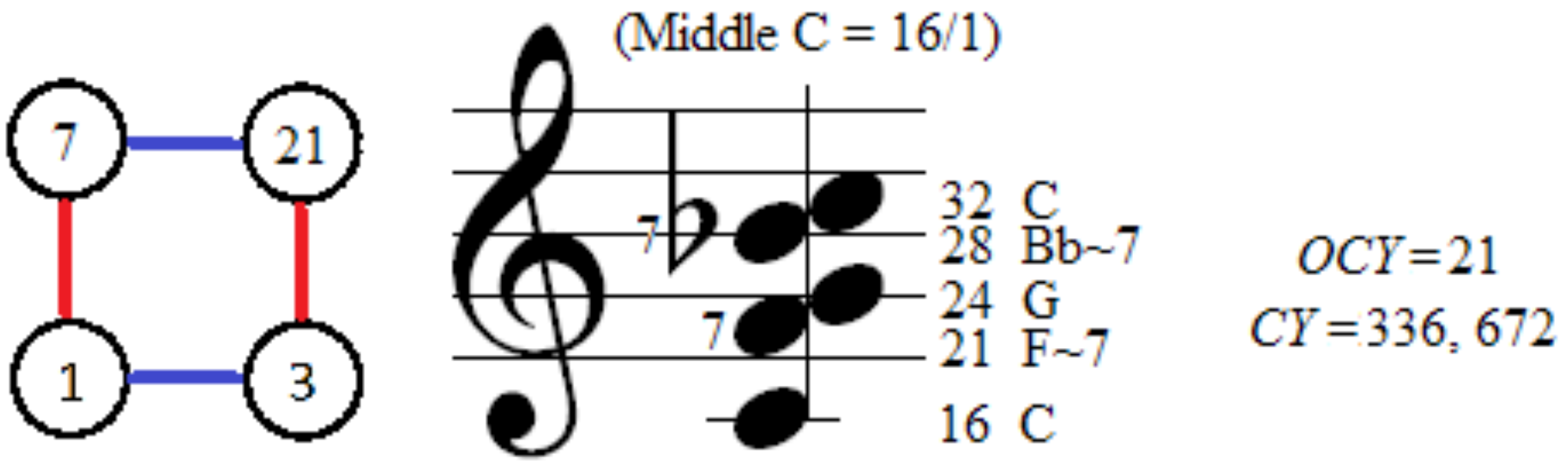

**Figure 9: Diagram of *OddComplexitySpace* lattice for the number 21, with scale**

The case *OCY* = 21 is illustrated in Figure 9. Here there is no perfect fifth due to the lack of a factor of 5. In general, when comparing *CY* values, it is best to have the same number of notes in the scale. *OCY* = 15, 21, 27 all give four note scales, and have *mCY* values respectively of 240, 336 and 864. This shows that *OCY* = 21 gives a scale only slightly less consonant than *OCY* = 15, and much more consonant than OCY = 27 with only 1 prime factor.

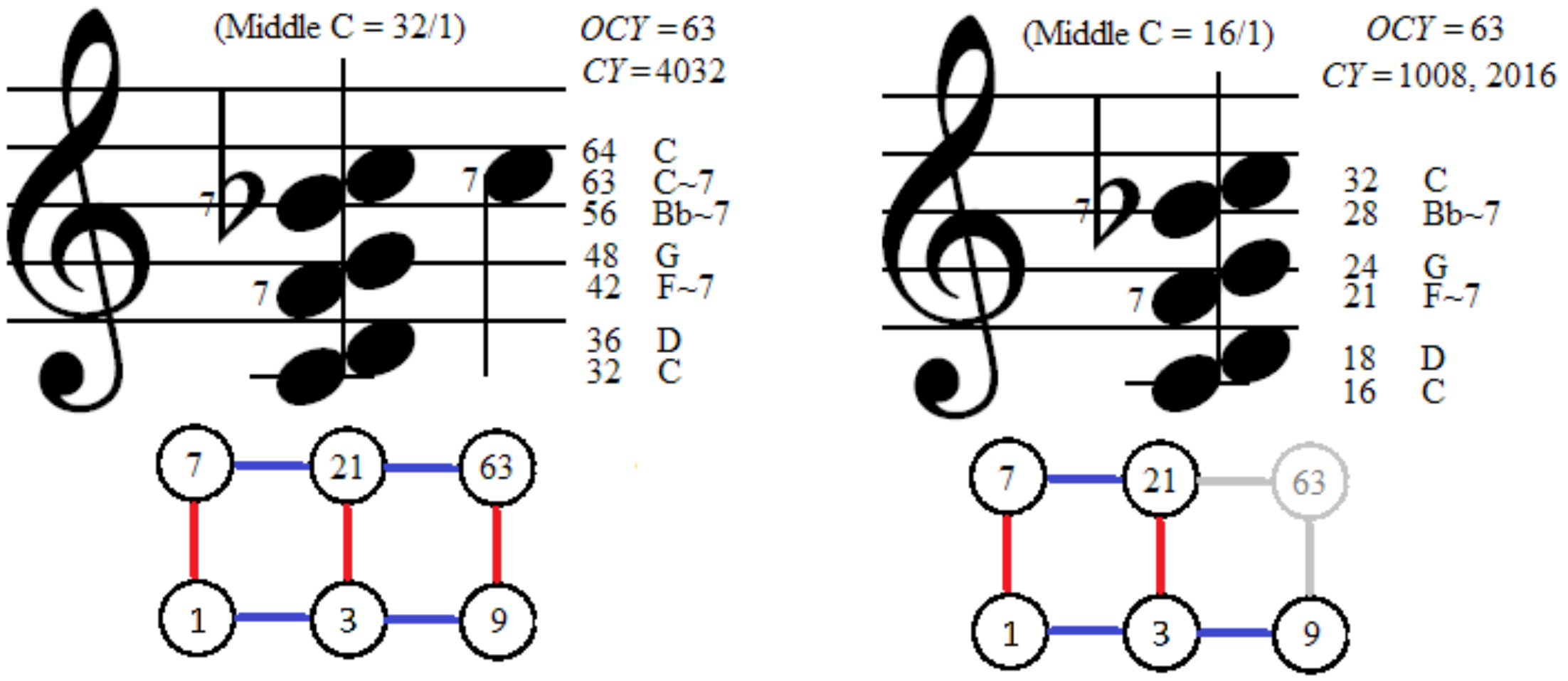

**Figure 10: Diagram of *OddComplexitySpace* lattice for the number 63: a) complete; b) subset**

In Figure 10a the value *OCY* = 63 is demonstrated, which extends the case of *OCY* = 21 with the new divisors 9 and 63. This has quite a high *CY* value of 4032 which is entirely due to the clash between 63 (C~7) and 64 (C) notes. However in Figure 10b the clashing divisor 63 has been removed, leaving dividors of (1, 3, 7, 9, 21) and a scale of (16, 18, 21, 24, 28, 32). This reduces the minimum *Complexity* value (call this *mCY*) from 4032 to 1008, which is due to the lower powers of 2 needed to bring all the odd numbers into the same octave. This results in smaller whole numbers too – 16 to 32, instead of 32 to 64.

Such an improvement in *CY* performance would also be obtained from removing the divisor 1 from the divisor lattice. However, removing any of 3, 9, 7, 21 has no effect on *CY*. For fixed *OCY*, the only thing affecting *CY* is the highest power of 2 necessary to bring all the notes into one octave. This power of 2 can be lowered by removing either the smallest divisor, or the largest, or both. In other words, by choosing a subset of divisors with comparable magnitude, *CY* will be closer to *OCY*.

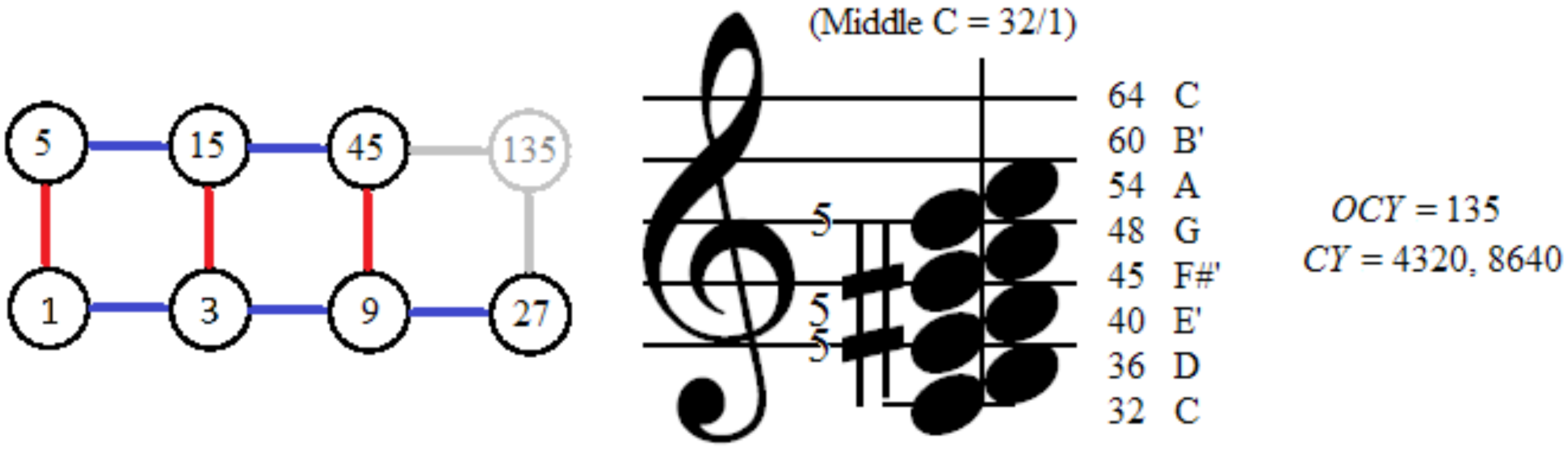

**Figure 11: Diagram of *OddComplexitySpace* lattice for the number 135, with a subset scale**

The lattice for *OCY* = 15 has been extended in Figure 11 to the lattice for *OCY* = 135. The factor 135 itself has been omitted to give a subset scale with seven notes and *mCY* = 4320. This scale is the seven-note diatonic major scale starting at G (the odd divisor 3), the most common scale in Western music. Alternatively, by removing the divisor 1 instead of 135, a natural minor scale is formed based on the note B' which has the same *OCY* and *CY* values as the major scale. In fact, among all seven-note scales, the major and natural minor scales have some of the lowest *OCY* and *CY* values, both with *mCY* = 4320. The only lower *mCY* values found (so far!) by the author for seven-note scales have *mCY* = 3360 and 3600. These scales are described below in this section. In any case, the diatonic major scale appears to be nearly optimal in *mCY*. (This could be no doubt be proved with a suitable search over all possible *OCY* values, see the section below on algorithms.) This is a clear case where mathematical efficiency and aesthetic beauty coincide, giving a reason to investigate and classify the statistical properties of all possible harmonies, to help discover those which may have the greatest aesthetic merit.

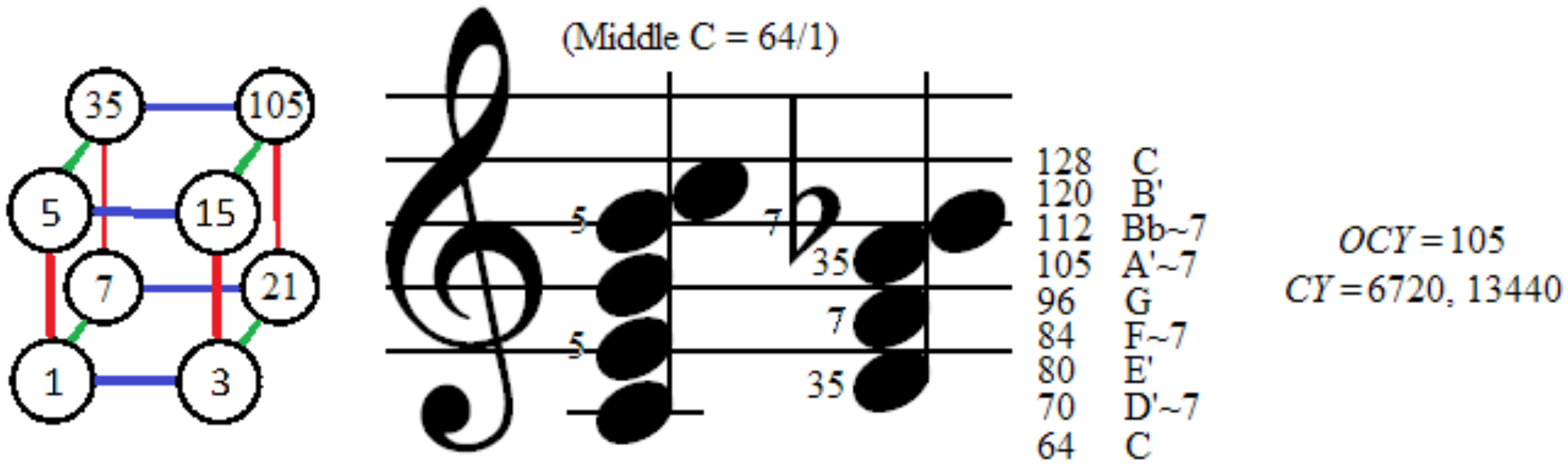

**Figure 12: Diagram of *OddComplexitySpace* lattice for the number 105, with a scale**

*OCY* = 105 is an interesting case since 105 is the first odd number with three distinct prime factors; 3, 5 and 7. This means the scale contains both perfect fifths (3), major thirds (5) and dominant seventh (7) notes. It has *mCY* = 6720, which can be reduced by removing factors 1 and/or 105 from the scale.

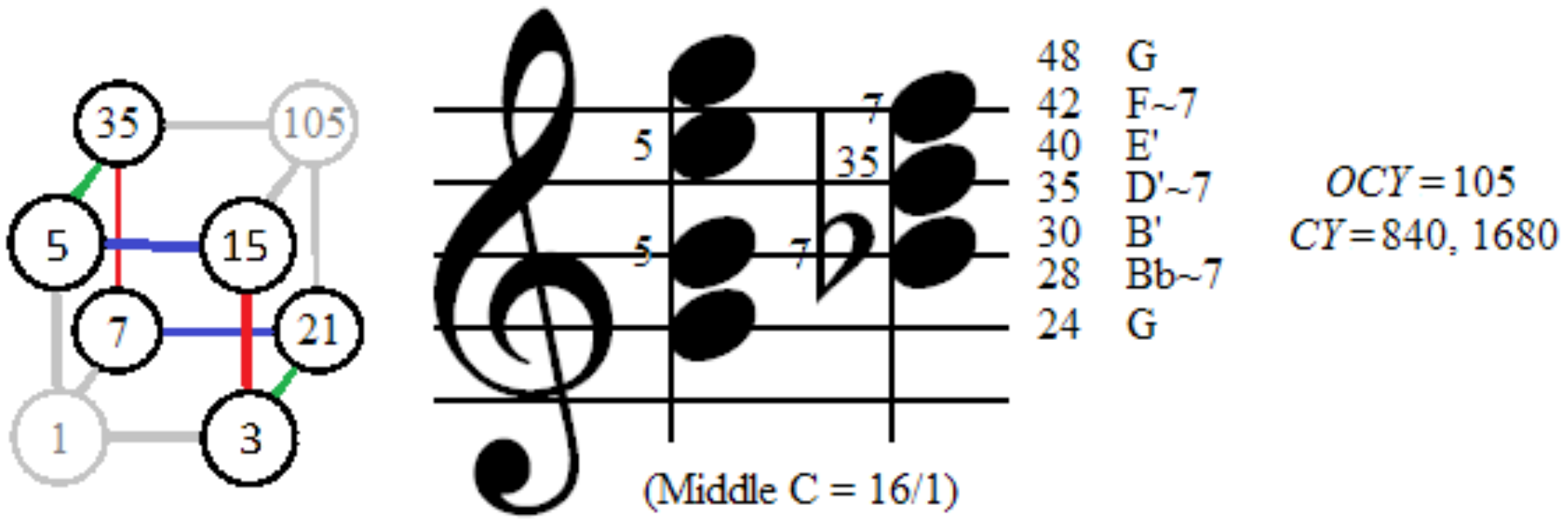

**Figure 13: Diagram of *OddComplexitySpace* lattice for the number 105, with a subset scale**

In Figure 13 the two outside factors 1, 105 have been removed from the divisor lattice for *OCY* = 105 to give a six-note scale with *mCY* = 840, a big improvement on 6720 for the eight-note scale. Moreover, either of the seven-note scales (from removing either 1 or 105) have *mCY* = 3360, which is lower than the value of 4320 for the diatonic major scale. The harmony here is also richer than for the major scale, due to the factors of 7 in many intervals. This scale would likely be interesting to compose with. A clue to how to compose with it is the embedding in the Tonnetz, which allows the 'shape' of each chord to be visualised. Two reasons why the major scale may currently be more popular than this scale is that the major scale is simpler, and the major scale is more even, having a *MaxRatio* of 9/8, whereas this scale has a *MaxRatio* of 7/6.

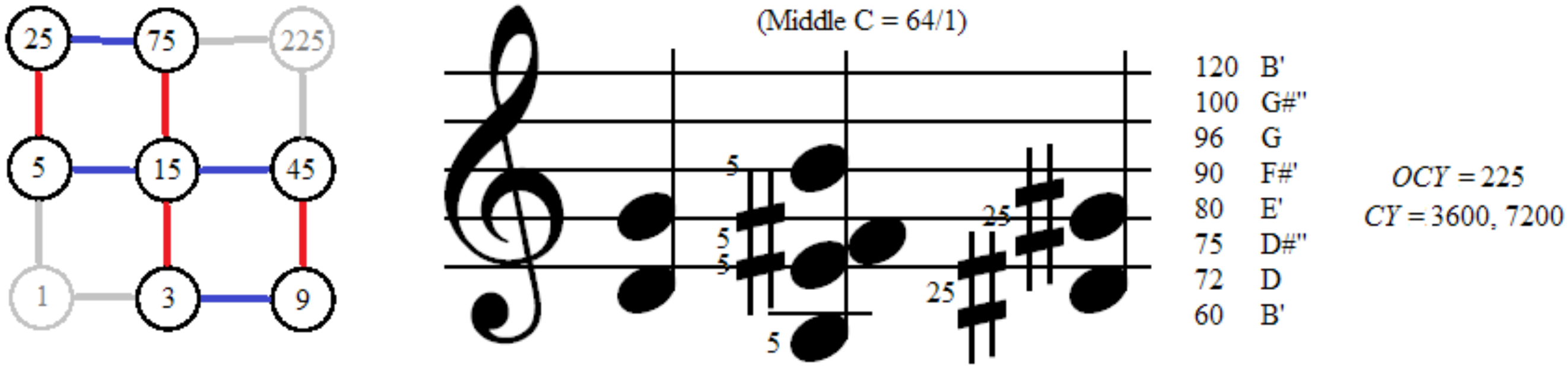

**Figure 14: Diagram of *OddComplexitySpace* lattice for the number 225, with a subset scale**

In Figure 14 a lattice for *OCY* = 225 has been given, which has two factors of 3 and two factors of 5. The lowest (1) and highest (225) divisors have been removed, giving a seven-note scale with *mCY* = 3600, again lower than the value of 4320 for the major scale. This scale is formed of all the major and minor triads for the note B' its central note. Reasons why this consonant scale might not be more popular than the major scale include: it is more uneven (*MaxRatio* of 6/5) and every triad has a B' in it, giving it less variety in triads and less supporting of chord progression than the major scale, which allows (for example) progression from F major to C major to G major. However, this scale should still be interesting to compose with due to its consonance.

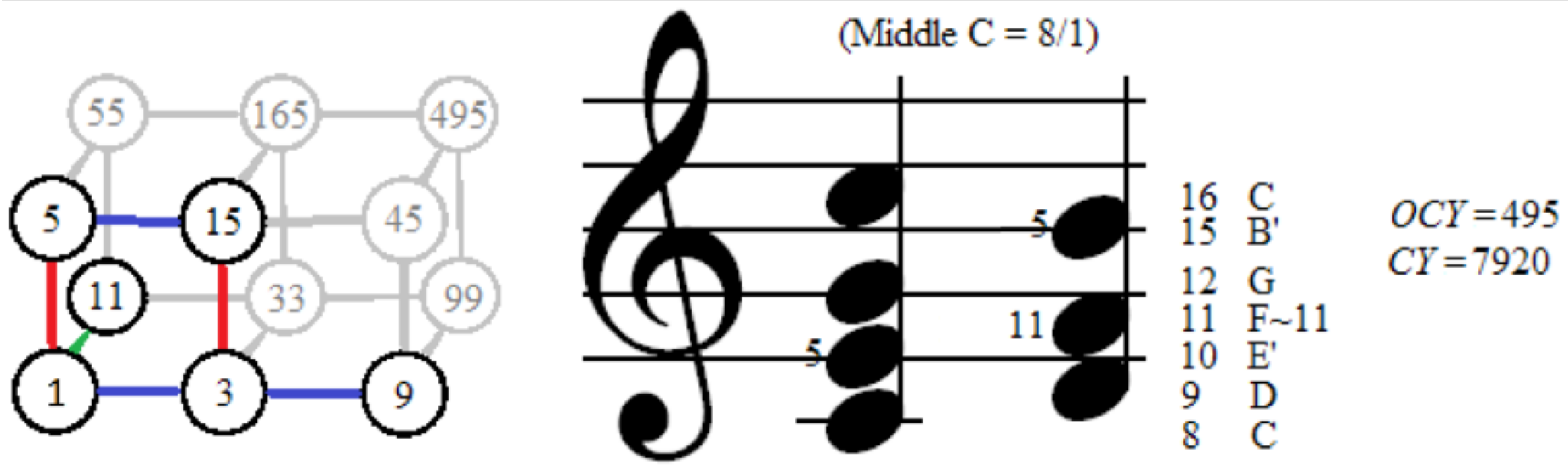

**Figure 15: Diagram of *OddComplexitySpace* lattice for the number 495, with a subset scale**

A final example (which is not intended to be highly consonant) introduces 11$^{th}$ harmonics into a scale (see Ryan 2016 in which notation F~11 represents the pitch class of 11/8). In Figure 15 the lattice for *OCY* = 495 is given, with only the divisors 15 and below, leading to a six-note scale. The value *mCY* = 7920 is higher than for the seven-note scales above, leading to the conclusion that 11$^{th}$ harmonics may not be helpful to maximise scale consonance. It is probably the case that as scale size increases, the scales with lowest *mCY* need to contain higher and higher primes. This conjecture is certainly worth investigating, since earlier on it was demonstrated for a scale with four distinct pitch classes that two primes (*OCY* = 15) outperformed one prime (*OCY* = 27) in terms of producing consonant scales with low *mCY* value.

## 15) Scales which are approximately equally spaced from interval splitting

As seen above, every odd number produces an octave based scale through multiplying its divisors by powers of 2. A problem is that these scales are not usually evenly spread. For example, 15 produces the divisors (1, 3, 5, 15) and thus the scale (8, 10, 12, 15, 16) in which the *MinRatio* of 16/15 is much smaller than the *MaxRatio* of 5/4. It would be good to have a procedure to produce whole-numbered scales where the *MinRatio* and *MaxRatio* were as close as possible. (Note this is not simply equal-tempering, since equal-tempering loses all the whole-number consonance from non-octave intervals, whereas this procedure will retain whole numbers and consonance, maximising the consonance if possible.)

The suggested procedure is as follows. Start with the base interval for the scale (normally an octave, 2/1; for Bohlen-Pierce scales it is 3/1). Subsequently, evenly divide the interval using whole numbers and equal increments. Then subdivide evenly one or more times again, with the aim to get the end frequency ratios as close as possible. For example: an octave 1:2 might be divided into 2:3:4 or 3:4:5:6, all of the increments are 1; then (selecting 2:3:4) the ratio 2:3 could be divided three times into 6:7:8:9, and 3:4 could be divided twice into 6:7:8; since the whole numbers in 6:7:8:9 and 6:7:8 are roughly equal, this makes the frequency ratios as nearly equal as possible. Then the final scale is obtained from stitching these compound ratios together, usually in the same order as the original splitting: this gives a final scale of 12:14:16:18:21:24 which the reader can verify has *MaxRatio* and *MinRatio* close together (7/6 and 9/8 respectively).

**Table 10: Results of splitting some consecutive intervals**

| 1, 2 | 2, 3 | 3, 4 | 4, 5 | 5, 6 |
|---|---|---|---|---|
| 2, 3, 4 | 4, 5, 6 | 6, 7, 8 | 8, 9, 10 | 10, 11, 12 |
| 3, 4, 5, 6 | 6, 7, 8, 9 | 9, 10, 11, 12 | 12, 13, 14, 15 | 15, 16, 17, 18 |
| 4, 5, 6, 7, 8 | 8, 9, 10, 11, 12 | 12, 13, 14, 15, 16 | 16, 17, 18, 19, 20 | 20, 21, 22, 23, 24 |
| 5, 6, 7, 8, 9, 10 | 10, 11, 12, 13, 14, 15 | 15, 16, 17, 18, 19, 20 | 20, 21, 22, 23, 24, 25 | 25, 26, 27, 28, 29, 30 |
| 6, 7, 8, 9, 10, 11, 12 | 12, 13, 14, 15, 16, 17, 18 | 18, 19, 20, 21, 22, 23, 24 | 24, 25, 26, 27, 28, 29, 30 | 30, 31, 32, 33, 34, 35, 36 |

The raw data needed for this procedure are the ratios obtained from even splitting of various intervals. Some of these are given in Table 10. The first column enumerates the possibilities when evenly splitting an octave. The two simplest options are 2:3:4 and 3:4:5:6. Hence the next four columns of Table 10 give the possibilities for splitting the ratios 2:3, 3:4, 4:5 and 5:6. The octave is iteratively split when either 2:3:4 is split (select one entry from 2nd column, and one entry from 3rd column), or alternatively 3:4:5:6 is split (select entries from 3rd, 4th and 5th columns).

Not all possibilities are evenly split; if splitting 2:3:4, then splitting 2:3 to 4:5:6 and 3:4 to 9:10:11:12 will not give an even split of the octave, since 5/4 is much bigger than 12/11. The key is to make the final integers approximately the same size.

Once a splitting of an octave has been obtained, the individual intervals could be rearrange to give another scale that is just as evenly split. However this tends to increase the *Complexity* value. The reason is because higher prime powers tend to occur. For example, in the 12:14:16:18:21:24 scale, the intervals are all 7/6, 8/7 or 9/8. Were two 7/6 intervals rearranged to be next to each other, a 49/36 interval would occur in the scale, and this would increase the power of 7 in *CY* from $7^1$ to $7^2$, making *CY* bigger. Hence the original order, the natural reorderings, or inversion ($f \to 1/f$) tend to give the most consonant results.

**Table 11: Results of splitting some non-consecutive intervals**

| 1, 4 | 1, 3 | 2, 5 | 5, 8 | 3, 5 | 5, 7 | 7, 9 |
|---|---|---|---|---|---|---|
| 2, 5, 8 | 2, 4, 6 =1, 2, 3 | 4, 7, 10 | 10, 13, 16 | 6, 8, 10 =3, 4, 5 | 10, 12, 14 =5, 6, 7 | 14, 16, 18 =7, 8, 9 |
| 3, 6, 9, 12 =1, 2, 3, 4 | 3, 5, 7, 9 | 6, 9, 12, 15 =2, 3, 4, 5 | 15, 18, 21, 24 =5, 6, 7, 8 | 9, 11, 13, 15 | 15, 17, 19, 21 | 21, 23, 25, 27 |
| 4, 7, 10, 13, 16 | 4, 6, 8, 10, 12 =2, 3, 4, 5, 6 | 8, 11, 14, 17, 20 | 20, 23, 26, 29, 32 | 12, 14, 16, 18, 20 =6, 7, 8, 9, 10 | 20, 22, 24, 26, 28 =10, 11, 12, 13, 14 | 28, 30, 32, 34, 36 =14, 15, 16, 17, 18 |

As mentioned above, we can start with intervals other than the octave. In Table 11 the double octave (1:4) and the Bohlen-Pierce tritave (1:3) have both been split in the 1st and 2nd columns respectively. From these, the first results are 1:4 splitting to 2:5:8, and 1:3 splitting to 3:5:7:9 (ignoring 2:4:6 which simplifies to intervals already split above in Table 10). Then in the 3rd and 4th columns 2:5 and 5:8 are split, and in the 5th, 6th and 7th columns 3:5, 5:7 and 7:9 are split.

As before, by judicious selection of entries from the correct columns, evenly split scales can be identified. However the evenness is not as good as in Table 10 (e.g. 1:4 splits to 2:5:8, and 2:5 is a much larger interval than 5:8). This can be compensated for by splitting the larger interval into more sub-segments than the smaller interval. (for 1:4) the intervals. Overall, splitting the octave is likely to be more useful than splitting other intervals.

**Table 12: Four examples of evenly split scales**

| **Original interval** | 1, 2 | 1, 2 | 1, 2 | 1, 4 |
|---|---|---|---|---|
| **Split interval** | 2, 3, 4 | 2, 3, 4 | 3, 4, 5, 6 | 1, 3, 4 |
| **Sequence (1)** | 2, 3 then 3, 4 | 2, 3 then 3, 4 | 3, 4 then 4, 5<br>then 5, 6 | 1, 3 then 3, 4<br>=<br>3, 5, 7, 9 then 6, 8 |
| **Sequence (2)** | 6, 7, 8, 9<br>then<br>6, 7, 8 | 8, 9, 10, 11, 12<br>9, 10, 11, 12 | 15, 16, 17, 18, 19, 20<br>16, 17, 18, 19, 20<br>15, 16, 17, 18 | 6, 7, 8, 9, 10<br>15, 17, 19, 21<br>7, 8, 9<br>6, 7, 8 |
| **SOLVE** | (Piece | together | the | sub-intervals) |
| **Fractions for scale** | 1/1, 7/6, 4/3, 3/2, 7/4, 2/1 | 1/1, 9/8, 5/4, 11/8, 3/2, 5/3, 11/6, 2/1 | 1/1, 16/15, 17/15, 6/5, 19/15, 4/3, 17/12, 3/2, 19/12, 5/3, 16/9, 17/9, 2/1 | 1/1, 7/6, 4/3, 3/2, 5/3, 17/9, 19/9, 7/3, 8/3, 3/1, 7/2, 4/1 |
| **Integers for scale** | 12, 14, 16, 18, 21, 24 | 24, 27, 30, 33, 36, 40, 44, 48 | 180, 192, 204, 216, 228, 240, 255, 270, 285, 300, 320, 340, 360 | 18, 21, 24, 27, 30, 34, 38, 42, 48, 54, 63, 72 |
| **Scale description** | 5 notes per octave (Pentatonic scale) | 7 notes per octave | 12 notes per octave | 11 notes in 2 octaves (8 pitch classes unevenly spread) |
| ***CY (Complexity)*** | $1008 = 2^4 \cdot 3^2 \cdot 7$ | $23\,760 = 2^4 \cdot 3^3 \cdot 5 \cdot 11$ | $13\,953\,600 = 2^6 \cdot 3^3 \cdot 5^2 \cdot 17 \cdot 19$ | $4\,883\,760 = 2^4 \cdot 3^3 \cdot 5 \cdot 7 \cdot 17 \cdot 19$ |
| ***OCY (OddComplexity)*** | $63 = 3^2 \cdot 7$ | $1485 = 3^3 \cdot 5 \cdot 11$ | $218\,025 = 3^3 \cdot 5^2 \cdot 17 \cdot 19$ | $305\,235 = 3^3 \cdot 5 \cdot 7 \cdot 17 \cdot 19$ |
| ***MNR (MinRatio)*** | 203 cents<br>9/8 | 151 cents<br>12/11 | 89 cents<br>20/19 | 173 cents<br>21/19 |
| ***MXR (MaxRatio)*** | 267 cents<br>7/6 | 203 cents<br>9/8 | 112 cents<br>16/15 | 267 cents<br>7/6 |

Worked examples are given in Table 12 for three splittings of the octave 1:2, and one splitting of the double octave 1:4. These result in octave scales with 5, 7 and 12 notes, and a double octave scale with 5.5 notes per octave. Judging by the closeness of *MinRatio* and *MaxRatio*, this procedure has resulted in much more evenly spread scales than those produced by some sets of divisors (e.g. divisors of 15), without resulting in too high a *CY* value from having too many different prime factors.

**Table 13: Some scales with uneven increment but approximately equal intervals**

| **Description** | Diminished chord scale | Augmented chord scale |
|---|---|---|
| **Original interval** | 10, 12, 14, 17, 20 | 12, 15, 19, 24 |
| **Fractions for scale** | 1/1, 6/5, 7/5, 17/10, 2/1 | 1/1, 5/4, 19/12, 2/1 |
| ***CY (Complexity)*** | $7140 = 2^2 \cdot 3 \cdot 5 \cdot 7 \cdot 17$ | $2280 = 2^3 \cdot 3 \cdot 5 \cdot 19$ |
| ***OCY (OddComplexity)*** | $1785 = 3 \cdot 5 \cdot 7 \cdot 17$ | $285 = 3 \cdot 5 \cdot 19$ |
| ***MNR (MinRatio)*** | 267 cents – 7/6 | 386 cents – 5/4 |
| ***MXR (MaxRatio)*** | 336 cents – 17/14 | 409 cents – 19/15 |

The procedure for splitting octaves evenly is not the only way of producing scales with roughly equal intervals. In Table 13 a diminished chord scale and an augmented chord scale are given, and neither can be produced by evenly splitting an octave. The reason is that for both scales the increments are uneven. Although these scales are not as consonant as those from major triads, they still have interest, and probably the best way of producing them is to perform a computer search for approximations of 3-EDO and 4-EDO scales using whole numbers. Further work is needed to identify the best way of producing scales with *MinRatio* and *MaxRatio* as close as possible, whilst minimising *Complexity* and maximising consonance.

## 16) Scale size, Lattice shapes and Oddly Highly Composite Numbers

There is a link between 'oddly highly composite numbers' (OHCN) and the appearance of scales with *N* pitch classes and minimal *OddComplexity*. A highly composite number (HCN) is a number with more divisors than any smaller number. An OHCN is an odd number with more divisors than any smaller odd number. The sequences of OHCNs is listed in an online database (Sloane 2016). Here are the first few terms, where *n* is the OHCN and *d*(*n*) is its (record) number of divisors for an odd number up to that point:

**Table 14:The first few oddly highly composite numbers *n* where *d*(*n*) is greater than all lower numbers**

| ***n* (OHCN)** | 1 | 3 | 9 | 15 | 45 | 105 | 225 | 315 |
|---|---|---|---|---|---|---|---|---|
| ***d*(*n*) divisor count** | 1 | 2 | 3 | 4 | 6 | 8 | 9 | 12 |
| **Prime factorisation of *n*** | 1 | 3 | $3^2$ | $3{\cdot}5$ | $3^2{\cdot}5$ | $3{\cdot}5{\cdot}7$ | $3^2{\cdot}5^2$ | $3^2{\cdot}5{\cdot}7$ $=2^0{\cdot}3^2{\cdot}5^1{\cdot}7^1$ |
| **Prime exponent vector (PEV)** | (0, 0, 0, 0) | (0, 1, 0, 0) | (0, 2, 0, 0) | (0, 1, 1, 0) | (0, 2, 1, 0) | (0, 1, 1, 1) | (0, 2, 2, 0) | (0, 2, 1, 1) |
| **Set of divisors** | 1 | 1, 3 | 1, 3, 9 | 1, 3, 5, 15 | 1, 3, 5, 9, 15, 45 | 1, 3, 5, 7, 15, 21, 35, 105 | 1, 3, 5, 9, 15, 25, 45, 75, 225 | 1, 3, 5, 7, 9, 15, 21, 35, 45, 63, 105, 315 |
| **Prime divisors** | none | 3 | 3 | 3, 5 | 3, 5 | 3, 5, 7 | 3, 5 | 3, 5, 7 |
| **Lattice dimensions** | 0 | 1 | 1 | 2 | 2 | 3 | 2 | 3 |
| **Lattice size** | 1 | 2 | 3 | 2×2 | 3×2 | 2×2×2 | 3×3 | 3×2×2 |

In Table 14 the OHCNs up to 315 are presented. Above 15 each OHCN is has two or more primes and so is composite, giving a divisor lattice with at least two directions. From 315 onwards the power of 3 is higher than any other power, giving more notes in the ‘3’ direction than in the directions for primes 5 and above. Examining the prime exponent vectors (PEVs – which represent the power of primes 2, 3, 5, 7… in the prime factorisation of *n* – demonstrated in factorisation of 315 above) and ignoring 2 (which always has zero power for odd numbers), the powers for 3, 5, 7… are in non-increasing order. This is since if the powers increased, say if 5 had a higher power in the PEV than 3 did, then in the factorisation of *n* a power of the higher prime 5 could be swapped for a power of the lower prime 3 giving a lower odd number $n' < n$ with the same number of divisors as *n*, which would be a contradiction since *n* was a OHCN which should have more divisors than any odd number smaller than it.

The reason OHCNs are relevant to music theory is that to produce an (octave-based) scale with *N* distinct pitch classes, an odd divisor lattice is required with at least *N* divisors. The lowest *OddComplexity* value where this occurs is the OHCN *n*, with *d*(*n*) divisors, where $N \leq d(n)$ for the first time. Examples of this are: to produce a scale with 4 notes, $15 \leq OCY$ since $d(15) = 4$ and 15 is an OHCN. Then to produce a scale with 5 or 6 notes, $45 \leq OCY$ since $d(45) = 6$ and 45 is an OHCN. Also to produce a scale with 7 or 8 notes, $105 \leq OCY$ since $d(105) = 8$ and 105 is an OHCN. Hence oddly highly composite numbers give minimal *OddComplexity* values for scales of a certain size.

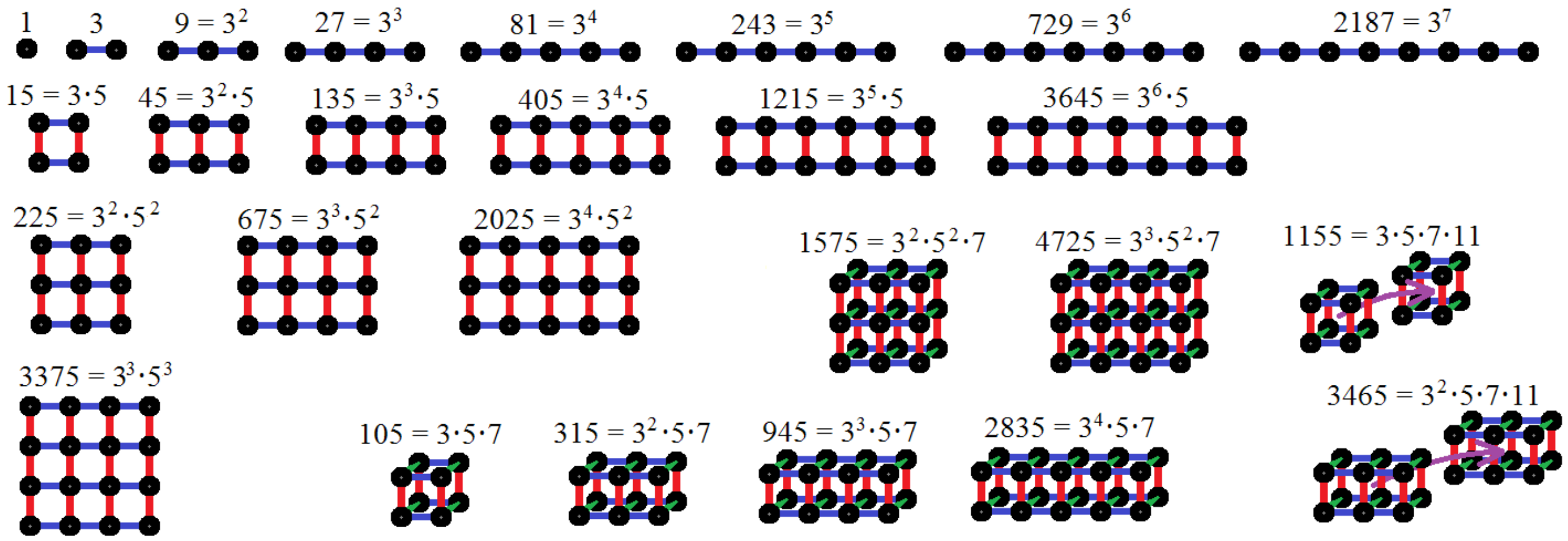

**Figure 16: Classification of odd divisor lattice shapes up to *OCY* = 2000.**
**Blue, red, green purple links correspond to multiplying by the primes 3, 5, 7, 11 respectively.**

In Figure 16 the shapes for odd divisor lattices have been classified up to a limit of *OCY* = 2000. The powers of each prime (PEV) are zero for the prime 2, and then non-increasing from the prime 3 onwards, in order to find the first time each lattice shape appears. For example, 21 is not a lattice shape in Figure 16 since it has the same shape of divisor lattice as 15, which is a smaller number; moreover $21 = 2^0 \cdot 3^1 \cdot 5^0 \cdot 7^1$ so its PEV of (0, 1, 0, 1) is not non-increasing from prime 3 onwards; however the PEV of $15 = 2^0 \cdot 3^1 \cdot 5^1 \cdot 7^0$ is (0, 1, 1, 0) which is indeed non-increasing from 3 onwards. In the diagram, some of the lattices have *OCY* > 2000; these have been included to demonstrate the set of *OCY* ≤ 2000 has been completed. Hence the unique odd divisor lattice shapes below *OCY* = 2000 appear first at the following values of *OCY*: 1, 3, 9, 15, 27, 45, 81, 105, 135, 225, 243, 315, 405, 675, 729, 945, 1155, 1215 and 1575 (see OEIS 2016, sequence 'A147516' regarding 'prime signatures'). Note that all OHCNs will give a new lattice shape, and so all OHCNs will be in this list; however not vice versa since not all new lattice shapes have more divisors than lower numbers. An example is 81 with five divisors which gives a new lattice shape, but 45 < 81 and 45 is an OHCN giving the first lattice shape with six divisors.

This leads on to an outline of an algorithm to identify new scales of length *N*, up to limiting values of *OCY* and *mCY*:

1. Find the new divisor lattice shapes with at least *N* divisors. These give minimal *OCY* values for each lattice.
2. Each lattice can also give larger *OCY* values by changing each prime for a larger, non-minimal prime. For example $45 = 3^2 \cdot 5$ is minimal, and $175 = 5^2 \cdot 7$ is a larger *OCY* value obtained by changing the primes. For each lattice type, find these up to the specified limit on *OCY*.
3. For each lattice, and each *OCY* value *n*, there are $d(n) \geq N$ divisors. Loop over all subsets of size *N*; make an octave-based scale for each subset, calculate *OCY* value, calculate *CY* values across all reorderings, add the scale and its supporting *OCY* and *mCY* values to the set of stored chords.
4. Repeat this for all subsets of all *OCY* values for all lattices. Sort the final stored chord list by increasing *mCY* value. The scales with minimal *mCY* are identified at the top of the list.

Alternatively, to identify the candidate *OCY* values it may be simpler to loop over odd numbers up to the *OCY* limit, reject any odd number with too few divisors, and then loop over divisor subsets for each odd number. This modified algorithm is considered in the next section, and avoids a manual (or complex to implement) process of switching primes. Nonetheless, being aware of the shapes of divisor lattices is helpful to understand why some *OCY* values occur more often for scales of length *N*.

## 17) Outline of algorithm to identify scales with low *Complexity* value

The following algorithm in pseudo-code could be implemented to use invariant functions *OCY* and *CY* to help identify scales with high consonance, e.g. low *CY* values. The most optimal scales (lowest *CY*) for each scale length *N* would be expected to be most useful for scale-based composition in JI.

1. Start with $n = 1$. This variable *n* is the odd integer whose divisors will be turned into a scale. Looking for octave-based scales of length *N* with low *CY* value
2. Loop
    a. *n* –> *n*+2 (this takes *n* through all the odd numbers)
    b. Find the (odd) prime factorisation of *n*
    c. Use this to calculate *d*(*n*), the number of divisors of *n*
    d. Go back to step a. if $d(n) < N$ since then there are not enough divisors to make a length-*N* scale
    e. Otherwise there are '*d*(*n*) choose *N*' ways of choosing divisors of *n* to give a length-*N* scale. Most of these have $OCY = n$. Some smaller subsets may have $OCY < n$
    f. Loop over these subsets
        i. If any other acceptance conditions (of the form $MinRatio \geq 16/15$, $MaxRatio < 4/3$, etc), use them here. (If subset rejected, go to the next subset)
        ii. Also reject subsets with $OCY < n$ since they would have been found for earlier values of *n*.
        iii. Otherwise turn the odd numbers in the subset into an octave based scale using method from earlier, calculate *CY* values for all reorderings, calculate *mCY* value, add the scale plus its *OCY* and *mCY* values to the chord store
    g. End loop
    h. Exit outer loop if conditions hold, e.g. *n* too big, chord store is full, time expired, etc
    i. Otherwise repeat loop to get more chords
3. End loop
4. Chord store now holds a set of candidate chords. Sort chord store by increasing *mCY* value
5. At top of chord store we now have octave-based scales of length *N* with the smallest *mCY* values for the search conditions used, provided *n* was allowed to loop over all necessary values.

Comments on this algorithm include:

- A brute force approach (of checking all sets of *N* odd numbers started at 1, 3, 5… and incrementing the set each loop) tends to be wasteful, it identifies many chords with very high *CY* which are no use to us if optimising by *CY*. These are where the *N* notes don't have many prime factors in common. For example, to find low *mCY* scales with 4 distinct pitch classes, it is not efficient to check (10, 11, 14, 17, 20) with *CY* = 26 180, when a nearby scale (10, 12, 15, 18, 20) has *CY* = 180.

- Modifying the above algorithm to have *n* –> *n*+1 and taking divisor subsets of both odd and even numbers leads to subsets where different notes will be multiples of 2 apart, which are not then octave-based scales with *N* notes. This approach would be helpful for other tasks, for example identifying low *Complexity* chords with *TotalRatio* < 8 (3 octaves).
- For the original algorithm: given odd *n* = *OCY*, and given an overall ceiling on *CY*, this also gives a ceiling for $CY_2 = CY/OCY$. Then any divisors of *n* (one large, one small) whose ratio is greater than $CY_2$ cannot be chosen in the same subset. This allows some spread-out subsets to be automatically rejected. Moreover, this condition on $CY_2$ means that generally the best subsets are the ones which use consecutive divisors, or nearly consecutive divisors (say if other conditions on *MinRatio* or *MaxRatio* are present). Modifying the algorithm to only search subsets which will have the right $CY_2$ values will dramatically reduce the number of subsets searched, and provide a much more efficient algorithm.
- Subsets with the lowest $CY_2$ values are usually found near the 'middle' of the divisor lattice since the factors are closer spaced there. E.g. for 15 the factors are (1, 3, 5, 15) and on a log scale 3, 5 are much closer spaced than the outside ones. This is a general feature of divisor lattices, so we could expect octave based scales with low *mCY* to originate from near the middle (on a log-scale) of the factor set of an odd number.

As an example algorithm setup: pentatonic scales have 5 pitch classes. Suppose we already know that scale (8, 9, 10, 12, 15, 16) has *OCY* = 45 and *mCY* = 720; also that scale (12, 14, 16, 18, 21, 24) has *OCY* = 63 and *mCY* = 1008, which are higher values, but give a more evenly spread scale. To find out if any other near-optimal pentatonic scales exist, the above algorithm could be run for *N* = 5, *OCY* ≤ 500, *mCY* ≤ 2000. A partial approach to this is given in Appendix 5.

Searches for three-note chords on a perfect fifth, and four note chords on an octave have also been carried out, and are given in Appendix 3 and Appendix 4 respectively.

By development of these kinds of algorithms, the most consonant chords and scales can be found for any given set of rules. Musical harmony and composition would be aided by improved knowledge of optimal scale and chord structures outside the limited set currently receiving widespread usage.

## 18) Conclusions

Throughout history, the consonant sounds of music have been explained by the fundamental theorem of harmony, which states that when the ratios between frequencies or string lengths are small whole numbers then the harmony sounds better. In literature this had been measured by the Benedetti height for two notes, and the *LCM* and *ESF* for any size of chord. Complete chords, tone lattices, major/minor distinctions and otonal/utonal concepts were also found in literature.

The concept of 'invariant' function was introduced here, which is a function of a chord which does not change on key transposition. Such invariant functions measure different aspects of the internal structure of a chord, and many different functions exists. To derive invariant functions from chords, ratios are usually taken; either from one note to another (a local ratio) or from larger parts of the chord to the GCD of the chord (a global ratio). Taking ratios also divides out any units in the chord notes (e.g. Hz for frequencies) and the ratio is dimensionless, a necessary (but not sufficient) condition for invariance.

The *Complexity* (*CY*) function is basic in understanding the structure of a chord, and is the *LCM* value of the chord divided by the *GCD* value. The *CY* value has a set of divisors (denoted *ComplexitySpace*, *CYS*), and all chords (when divided by their *GCD*) are a subset of this set of divisors. *CYS* itself is a 'complete' chord according to Euler's definition, and a rectangular-shaped set of points inside a *p*-limit Tonnetz, where *p* is the highest prime dividing *CY*. The position of *Chord*/*GCD* inside *CYS* gives coefficients for both *Otonality* and *Utonality*, which are generalisations of major and minor. How spread-out or skewed the *Chord* is within *CYS* can also be measured.

*CYS* is a rectangular subset of the *p*-limit Tonnetz with *n* dimensions. *CYS* can be projected into the space spanned by one or more of these dimensions (a prime projection). An important projection is to discard all information about the prime 2, mapping a *Chord* onto its odd components. This gives an odd complexity measure *OCY*, which is useful for measuring harmony up to octave equivalence, e.g. for sets of pitch classes, or for octave based scales.

A *Chord* can be given a set of weights which can represent any of: multiplicities, amplitudes, loudnesses of the notes. Applications of weighted chord analysis include: measuring properties of chords where the notes are different amplitudes or played on multiple instruments; measuring properties of waveforms where higher harmonics are present; measuring properties of chords made by combining such waveforms. One open issue is that the *Complexity* function does not appear to be well defined on waveforms with an infinite harmonic series; a midpoint approach might yield better results, i.e. it might converge.

Given a set of conditions specified by invariant functions, it is possible to derive algorithms for finding all of the harmonies which meet those conditions. Different combinations of condition may require different algorithms to find the chords efficiently, since a brute force approach is impractical where the number of notes increases (for the author's computer and using Excel, this was for more than 3 or 4 notes). Classifying chords by *CY* or *OCY* seems a promising shortcut. An example of search conditions might be: $N = 3$ (three notes), $TotalRatio = 3/2$, 240¢ < *MinRatio* < 462¢ (where cents ¢ are $1200\times\log_2$ of the value), sort by ascending *CY* value; this search is carried out in Appendix 3 where the chords meeting the *MinRatio* condition are highlighted. Using invariant functions in this way to classify and search harmony seems the most promising way of finding the best new harmonies for use in musical composition, to move away from twelve notes, to breath new vigour and infinite variety into the harmony of new music.

## 19) Nomenclature and Abbreviations

(For a list of functions and their abbreviations, see Appendix 1)

| | |
|---|---|
| *N*-EDO | Equal Division of the Octave into *N* parts (e.g. 12-EDO, 19-EDO, etc) |
| ASCII | A character set for computers which contains all common keyboard characters |
| Coprime | Whole numbers where the biggest number dividing evenly into them is 1 |
| EDO | A tuning system splitting an octave into equal parts to make a scale |
| JI | Just Intonation |
| *p*-limit | Rational numbers containing only primes up to *p* in their prime factorisations |
| *q* odd limit | Rational numbers whose numerator and denominator contain odd factors only up to *q* |

## 21) Author contact details

| | | |
|---|---|---|
| ORC ID | http://orcid.org/0000-0002-4785-9766 | Academic profile |
| arXiv | https://arxiv.org/a/ryan_d_1 | Papers (pre-prints) |
| SoundCloud | https://soundcloud.com/daveryan23/tracks | JI music examples |
| LinkedIn | https://www.linkedin.com/in/davidryan59 | Professional page |
| Email: | david ryan 1998 @ hotmail.com | (remove spaces) |

# Appendix 1 Function reference table

**Table 15: Functions presented in this paper. Invariant functions are in black, non-invariant functions are in grey, which are those which change if all the frequencies are multiplied by a constant, i.e. if the base frequency changes.**

| Function Name | Abbreviated Name | Description of Function | Link to equation |
|---|---|---|---|
| *Chord* | *CH* | The main input: a set of note frequencies for a chord. Normally multiplied by a constant until they are all whole numbers. | |
| *LogChord* | *LCH* | Base-2 logarithms of *CH* | Equation 5 |
| *N* | *N* | Size of set *CH*; number of notes | |
| *Chord*(*n*) | *CH*(*n*) | Frequency of the *n*th note in *CH* (*n*=1..*N*) | |
| *LogChord*(*n*) | *LCH*(*n*) | Base-2 logarithm of *CH*(*n*) | Equation 18 |
| *GCD* | *GCD* | Greatest Common Divisor of *CH* | |
| *LCM* | *LCM* | Lowest Common Multiple of *CH* | |
| *LogGCD* | *LGCD* | Base-2 logarithm of GCD | Equation 6 |
| *LogLCM* | *LLCM* | Base-2 logarithm of LCM | Equation 7 |
| *BenedettiHeight* | *BH* | $a \cdot b$, for interval $a/b$ in lowest terms | Equation 1 |
| *TenneyHeight* | *TH* | $\log_2(BH)$ | Equation 3 |
| *KeesHeight* | *KH* | Highest odd number dividing *a* or *b*, in $a/b$ reduced frequency ratio | Equation 2 |
| *EulerSweetnessFunction* | *ESF* | Add ($p$-1) for each prime factor (repeats allowed) *p* of *LCM*, then add 1 | |
| *Complexity* | *CY* | Ratio of *LCM* to *GCD* | Equation 4 |
| (minimum) *Complexity* | *mCY* | For a scale: minimum *CY* value obtained across cyclic re-orderings of the scale | |
| *LogComplexity* | *LCY* | Base-2 logarithm of *CY* | Equation 8 |
| *ComplexitySpace* | *CYS* | Set of divisors of *CY*, an integer | |
| *LogComplexitySpace* | *LCYS* | Base-2 logarithms of *CYS* | |
| *LogMidpoint* | *LM* | The average value of *LCH* minus *LGCD*; the geometric midpoint of *LCH* in *LCYS* | Equation 9 |
| *Otonality* (coefficient) | *OTC* | The negation of *UTC*. High for some chords made of small integers. | Equation 10 |
| *Utonality* (coefficient) | *UTC* | A coefficient for the position of *LM* within the extreme values possible for it | Equation 10 |
| *SpreadCoeff* | *SPC* | A coefficient for describing how 'spread out' *CH* is in *LCS* | Equation 11 |
| *Skewness* | *SK* | The direction of the 'skew' of the notes in *CH* | Equation 12 |
| (chord) *Ratio*(*m*,*n*) | *CR*(*m*,*n*) | *CH*(*n*)/*CH*(*m*) | Equation 13 |
| (chord) *Ratio*(*k*) | *CR*(*k*) | Ratio between consecutive frequencies in *CH* | Equation 14 |
| *MinRatio* | *MNR* | Minimum of *CR* | Equation 15 |
| *MaxRatio* | *MXR* | Maximum of *CR* | Equation 16 |
| *TotalRatio* | *TR* | Ratio between highest and lowest notes in *CH* | Equation 17 |
| *LogRatio*(*m*,*n*) | *LCR*(*m*,*n*) | Base-2 logarithm of *CR*(*m*,*n*) | Equation 19 |
| *LogRatio*(*k*) | *LCR*(*k*) | Base-2 logarithm of *CR*(*k*) | Equation 20 |
| *LogMinRatio* | *LMNR* | Base-2 logarithm of *MNR* | Equation 21 |
| *LogMaxRatio* | *LMXR* | Base-2 logarithm of *MXR* | Equation 22 |
| *LogTotalRatio* | *LTR* | Base-2 logarithm of *TR* | Equation 23 |
| *MinRatioCoeff* | *MNRC* | Coefficient for position of *LMNR* within its possible extreme values | Equation 24 |
| *MaxRatioCoeff* | *MXRC* | Coefficient for position of *LMXR* | Equation 25 |
| *TotalRatioCoeff* | *TRC* | Coefficient for position of *LTR* | Equation 26 |
| *P* | *P* | Set of prime numbers to project over | |
| $GCD_P$ | $GCD_P$ | Value of *GCD* restricted to only primes within set *P* | |
| $LCM_P$ | $LCM_P$ | Value of *LCM* restricted to only primes within set *P* | |
| $Complexity_P$ | $CY_P$ | For a prime set *P*, Ratio of $LCM_P$ to $GCD_P$ | |
| $Complexity_p$ | $CY_p$ ($CY_2$, $CY_3$, etc) | For a specific prime number *p*, the factor $p^k$ in the prime factorisation of *CY* | |
| *OddComplexity* | *OCY* | $CY_P$ value for *P* the set of primes excluding 2 (the odd primes) | |
| *OddComplexitySpace* | *OCYS* | The set of divisors of *OCY* | |
| *BohlenPierceComplexity* | *BPCY* | $CY_P$ value for *P* the set of primes excluding 2 and 3 | |
| *Weight* | *WT* | Set of weights or amplitudes for chord *CH*; sets must be the same size (*N*) | |
| *Weight*(*n*) | *WT*(*n*) | The *n*th weight or amplitude in *WT* | |
| *SumWeight* | *SWT* | Sum of weights in *WT* | Equation 27 |
| *WeightedLogMidpoint* | *WLM* | The weighted average value of *LCH* in *LCYS* | Equation 28 |
| *WeightedOtonality* | *WOTC* | The negation of *WUTC* | Equation 29 |
| *WeightedUtonality* | *WUTC* | A coefficient for the position of *WLM* within *LCY* | Equation 29 |

Some of the conventions used for the abbreviations:

- *L*<*X*> means the logarithm of <*X*>
- *W*<*X*> means a weighted version of <*X*>
- <*X*>*C* means a coefficient based on position of <*X*> within its extreme values
- <*X*>(*n*) means the *n*th value of (multiple-valued) <*X*>

# Appendix 2 Functions evaluated for some example chords

**Table 16: Some chord examples using small integers with a selection of (non-weighted) functions calculated for each**

| *CH* | *Chord* | 2, 3, 4 | 3, 4, 5 | 4, 5, 6, 7 | 4, 5, 6, 7, 8 | 5, 6, 7, 9, 11 | 5, 6, 7, 9, 11, 13 | 12, 15, 20, 30 | 1, 30, 60 |
|---|---|---|---|---|---|---|---|---|---|
| ***N*** | ***N*** | 3 | 3 | 4 | 5 | 5 | 6 | 4 | 3 |
| ***CY*** | ***Complexity*** | 12 | 60 | 420 | 840 | 6930 | 90090 | 60 | 60 |
| $\boldsymbol{CY_2}$ | $\boldsymbol{Complexity_2}$ | 4 | 4 | 4 | 8 | 2 | 2 | 4 | 4 |
| ***OCY*** | ***OddComplexity*** | 3 | 15 | 105 | 105 | 3465 | 45045 | 15 | 15 |
| $\boldsymbol{CY_3}$ | $\boldsymbol{Complexity_3}$ | 3 | 3 | 3 | 3 | 9 | 9 | 3 | 3 |
| ***BPCY*** | ***BohlenPierce-Complexity*** | 1 | 5 | 35 | 35 | 385 | 5005 | 5 | 5 |
| ***OTC*** | ***Otonality*** | 0.442 | 1.000 | 0.885 | 0.794 | 0.917 | 0.952 | -0.831 | -0.661 |
| ***SPC*** | ***SpreadCoeff*** | 0.229 | 0.102 | 0.069 | 0.073 | 0.064 | 0.059 | 0.167 | 0.874 |
| ***MNR*** | ***MinRatio*** | 4/3 | 5/4 | 7/6 | 8/7 | 7/6 | 7/6 | 5/4 | 2/1 |
| ***MXR*** | ***MaxRatio*** | 3/2 | 4/3 | 5/4 | 5/4 | 9/7 | 9/7 | 3/2 | 30/1 |
| ***TR*** | ***TotalRatio*** | 2/1 | 5/3 | 7/4 | 2/1 | 11/5 | 13/5 | 5/2 | 60/1 |
| ***MNRC*** | ***MinRatioCoeff*** | 0.830 | 0.874 | 0.826 | 0.771 | 0.782 | 0.807 | 0.731 | 0.339 |
| ***MXRC*** | ***MaxRatioCoeff*** | 0.170 | 0.126 | 0.098 | 0.096 | 0.092 | 0.079 | 0.164 | 0.661 |
| ***TRC*** | ***TotalRatioCoeff*** | 0.279 | 0.125 | 0.093 | 0.103 | 0.089 | 0.084 | 0.224 | 1.000 |

**Table 17: Comparison of three different dominant seventh chords (c.f. Garrett 2013)**

| | **Chord Fractions** | 1/1, 5/4, 3/2, 16/9 | 1/1, 5/4, 3/2, 9/5 | 1/1, 5/4, 3/2, 7/4 |
|---|---|---|---|---|
| ***CH*** | ***Chord*** (lowest terms) | 36, 45, 54, 64 | 20, 25, 30, 36 | 4, 5, 6, 7 |
| | **Score** | 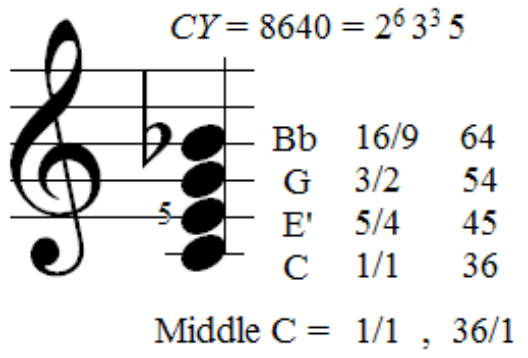  | 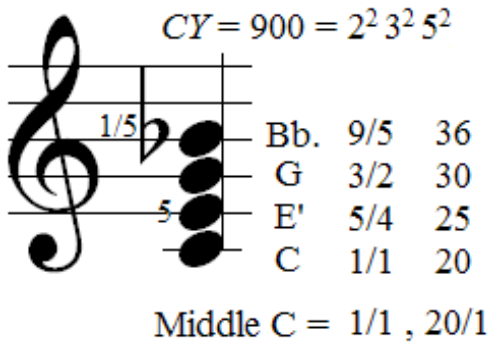  | 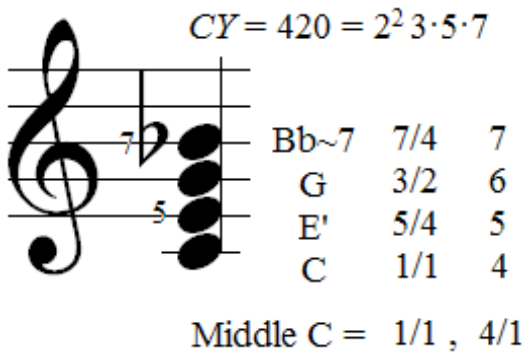  |
| ***N*** | ***N*** | 4 | 4 | 4 |
| ***CY*** | ***Complexity*** | 8640 | 900 | 420 |
| ***$CY_2$*** | ***$Complexity_2$*** | 64 | 4 | 4 |
| ***OCY*** | ***OddComplexity*** | 135 | 225 | 105 |
| ***$CY_3$*** | ***$Complexity_3$*** | 27 | 9 | 3 |
| ***BPCY*** | ***BohlenPierce-Complexity*** | 5 | 25 | 35 |
| ***OTC*** | ***Otonality*** | 0.2858 | 0.0596 | 0.8852 |
| ***SPC*** | ***SpreadCoeff*** | 0.0472 | 0.0640 | 0.0691 |
| ***MNR*** | ***MinRatio*** | 32/27 | 6/5 | 7/6 |
| ***MXR*** | ***MaxRatio*** | 5/4 | 5/4 | 5/4 |
| ***TR*** | ***TotalRatio*** | 16/9 | 9/5 | 7/4 |
| ***MNRC*** | ***MinRatioCoeff*** | 0.8859 | 0.9305 | 0.8264 |
| ***MXRC*** | ***MaxRatioCoeff*** | 0.0817 | 0.0695 | 0.0981 |
| ***TRC*** | ***TotalRatioCoeff*** | 0.0635 | 0.0864 | 0.0926 |

**Table 18: Comparison of three different ninth chords**

| | **Chord Name** | Diminished Ninth | Dominant Ninth | Augmented Ninth (c.f. Hendrix Chord) |
|---|---|---|---|---|
| ***CH*** | ***Chord*** | 8, 10, 12, 14, 17 | 8, 10, 12, 14, 18 | 8, 10, 12, 14, 19 |
| | **Score** | 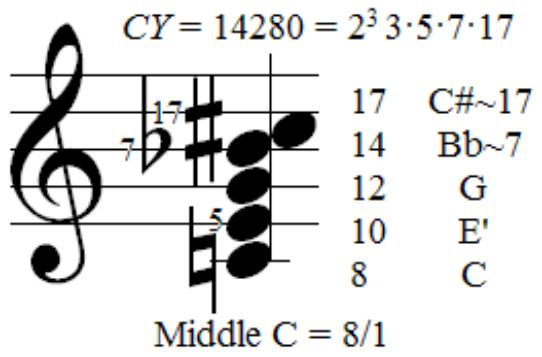 | 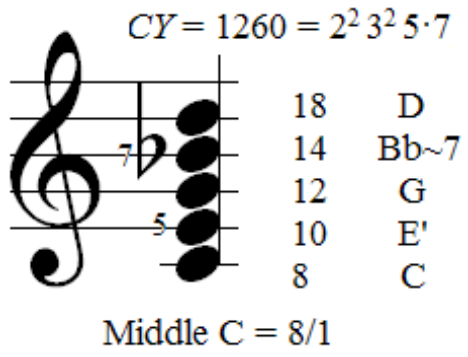 | 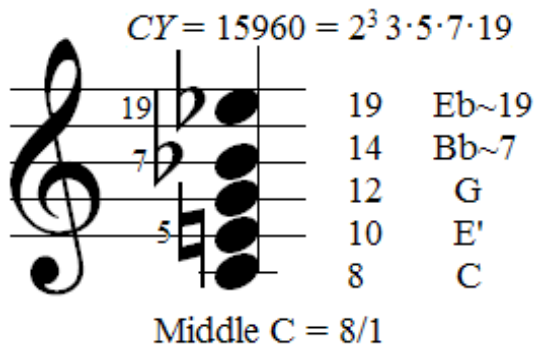 |
| ***N*** | ***N*** | 5 | 5 | 5 |
| ***CY*** | ***Complexity*** | 14 280 | 1260 | 15 960 |
| ***$CY_2$*** | ***$Complexity_2$*** | 8 | 4 | 8 |
| ***OCY*** | ***OddComplexity*** | 1785 | 315 | 1995 |
| ***$CY_3$*** | ***$Complexity_3$*** | 3 | 9 | 3 |
| ***BPCY*** | ***BohlenPierce-Complexity*** | 595 | 35 | 665 |
| ***OTC*** | ***Otonality*** | 0.8068 | 0.8327 | 0.8090 |
| ***SPC*** | ***SpreadCoeff*** | 0.0546 | 0.0778 | 0.0608 |
| ***MNR*** | ***MinRatio*** | 7/6 | 7/6 | 7/6 |
| ***MXR*** | ***MaxRatio*** | 5/4 | 9/7 | 19/14 |
| ***TR*** | ***TotalRatio*** | 17/8 | 9/4 | 19/8 |
| ***MNRC*** | ***MinRatioCoeff*** | 0.8180 | 0.7604 | 0.7128 |
| ***MXRC*** | ***MaxRatioCoeff*** | 0.0614 | 0.0799 | 0.1374 |
| ***TRC*** | ***TotalRatioCoeff*** | 0.0788 | 0.1136 | 0.0894 |

## Appendix 3 Classification of 3-note triads on perfect fifths with low *Complexity*

**Table 19: Each triad is of the form (2*k*, *m*, 3*k*) for some *m*, *k*, where 2*k* < *m* < 3*k*. The value 2*k* was searched up to 2*k*=138.**

| **2*k*** | ***m*** | **3*k*** | ***Complexity* *CY*** | ***Otonality* *OTC*** | ***Ratio*(1) *CR*(1)** | ***Ratio*(2) *CR*(2)** | ***Ratio*(1) in cents** |
|---|---|---|---|---|---|---|---|
| 4 | 5 | 6 | 60 | 0.6614 | 5/4 | 6/5 | 386.31 |
| 10 | 12 | 15 | 60 | -0.6614 | 6/5 | 5/4 | 315.64 |
| 6 | 8 | 9 | 72 | 0.1621 | 4/3 | 9/8 | 498.04 |
| 8 | 9 | 12 | 72 | -0.1621 | 9/8 | 4/3 | 203.91 |
| 6 | 7 | 9 | 126 | 0.5457 | 7/6 | 9/7 | 266.87 |
| 14 | 18 | 21 | 126 | -0.5457 | 9/7 | 7/6 | 435.08 |
| 10 | 14 | 15 | 210 | 0.1388 | 7/5 | 15/14 | 582.51 |
| 14 | 15 | 21 | 210 | -0.1388 | 15/14 | 7/5 | 119.44 |
| 8 | 11 | 12 | 264 | 0.5028 | 11/8 | 12/11 | 551.32 |
| 22 | 24 | 33 | 264 | -0.5028 | 12/11 | 11/8 | 150.64 |
| 10 | 11 | 15 | 330 | 0.4449 | 11/10 | 15/11 | 165.00 |
| 22 | 30 | 33 | 330 | -0.4449 | 15/11 | 11/10 | 536.95 |
| 14 | 16 | 21 | 336 | 0.0927 | 8/7 | 21/16 | 231.17 |
| 16 | 21 | 24 | 336 | -0.0927 | 21/16 | 8/7 | 470.78 |
| 10 | 13 | 15 | 390 | 0.4605 | 13/10 | 15/13 | 454.21 |
| 26 | 30 | 39 | 390 | -0.4605 | 15/13 | 13/10 | 247.74 |
| 14 | 20 | 21 | 420 | 0.1262 | 10/7 | 21/20 | 617.49 |
| 20 | 21 | 30 | 420 | -0.1262 | 21/20 | 10/7 | 84.47 |
| 12 | 13 | 18 | 468 | 0.4172 | 13/12 | 18/13 | 138.57 |
| 26 | 36 | 39 | 468 | -0.4172 | 18/13 | 13/12 | 563.38 |
| 18 | 20 | 27 | 540 | 0.0812 | 10/9 | 27/20 | 182.40 |
| 20 | 27 | 30 | 540 | -0.0812 | 27/20 | 10/9 | 519.55 |
| 18 | 22 | 27 | 594 | 0.0949 | 11/9 | 27/22 | 347.41 |
| 22 | 27 | 33 | 594 | -0.0949 | 27/22 | 11/9 | 354.55 |

This table classifies the triads of low *Complexity* which span a perfect fifth. The simplest triads are the standard major (4, 5, 6) and minor (10, 12, 15) triads which both have *Complexity* = 60. Usually for a triad the inner note should be more than about 240 cents away from either surrounding note, which would be *Ratio*(1) cents values of [240, 462]. Three other chords (highlighted) which satisfy this are (6, 7, 9) subminor triad (CY=126), (10, 13, 15) ultramajor triad (CY=390) and (18, 22, 27) neutral triad (CY=594). Between them, these four types of triads use the primes 5, 7, 13, 11 which are the four primes after 2 and 3 which describe the a perfect fifth of 2:3 frequency ratio.

## Appendix 4 Classification of 4-note chords with defined restrictions

**Table 20: Each chord is of the form (*k*, *m*, *n*, 2*k*) for some *m*, *n*, *k* where *k* < *m* < *n* < 2*k*. The list has been filtered to remove any *MinRatio* values below 240 cents. The value *k* was searched up to *k*=36.**

| *k* | *m* | *n* | 2*k* | *Complexity CY* | *OddComplexity OCY* | *BPCY* | *MinRatio MNR* | *MinRatio* (cents) |
|---|---|---|---|---|---|---|---|---|
| 3 | 4 | 5 | 6 | 60 | 15 | 5 | 6/5 | 315.64 |
| 10 | 12 | 15 | 20 | 60 | 15 | 5 | 6/5 | 315.64 |
| 4 | 5 | 6 | 8 | 120 | 15 | 5 | 6/5 | 315.64 |
| 5 | 6 | 8 | 10 | 120 | 15 | 5 | 6/5 | 315.64 |
| 12 | 15 | 20 | 24 | 120 | 15 | 5 | 6/5 | 315.64 |
| 15 | 20 | 24 | 30 | 120 | 15 | 5 | 6/5 | 315.64 |
| 5 | 6 | 7 | 10 | 210 | 105 | 35 | 7/6 | 266.87 |
| 21 | 30 | 35 | 42 | 210 | 105 | 35 | 7/6 | 266.87 |
| 6 | 7 | 9 | 12 | 252 | 63 | 7 | 7/6 | 266.87 |
| 7 | 9 | 12 | 14 | 252 | 63 | 7 | 7/6 | 266.87 |
| 9 | 12 | 14 | 18 | 252 | 63 | 7 | 7/6 | 266.87 |
| 14 | 18 | 21 | 28 | 252 | 63 | 7 | 7/6 | 266.87 |
| 18 | 21 | 28 | 36 | 252 | 63 | 7 | 7/6 | 266.87 |
| 21 | 28 | 36 | 42 | 252 | 63 | 7 | 7/6 | 266.87 |
| 6 | 7 | 10 | 12 | 420 | 105 | 35 | 7/6 | 266.87 |
| 7 | 10 | 12 | 14 | 420 | 105 | 35 | 7/6 | 266.87 |
| 30 | 35 | 42 | 60 | 420 | 105 | 35 | 7/6 | 266.87 |
| 35 | 42 | 60 | 70 | 420 | 105 | 35 | 7/6 | 266.87 |
| 15 | 18 | 25 | 30 | 450 | 225 | 25 | 6/5 | 315.64 |
| 10 | 13 | 15 | 20 | 780 | 195 | 65 | 15/13 | 247.74 |
| 13 | 15 | 20 | 26 | 780 | 195 | 65 | 15/13 | 247.74 |
| 15 | 20 | 26 | 30 | 780 | 195 | 65 | 15/13 | 247.74 |
| 26 | 30 | 39 | 52 | 780 | 195 | 65 | 15/13 | 247.74 |
| 30 | 39 | 52 | 60 | 780 | 195 | 65 | 15/13 | 247.74 |

The top and bottom notes of these chords are an octave apart, but the middle two notes can be any in-between integers which give *MinRatio* above 240 cents. Sorting by *Complexity* value shows that there are relatively few *Complexity* values possible, the first few of which are 60, 10, 210, 252, 420, 450, 780. Apart from the last value, these are all 7-smooth numbers. This indicates that when the number of notes in the chord increases, for low *Complexity* the chords increasingly tend to have only the lower prime numbers. This of course agrees with classical harmony which is based on the numbers 2, 3, 5 and sometimes 7, indicating a preference for low *Complexity* values in classical harmony.

## Appendix 5 Pentatonic scales with low *Complexity* value

A pentatonic scale is of the form $k_1 < k_2 < k_3 < k_4 < k_5 < 2k_1$ for some whole numbers $k_i$. A brute force search was conducted on $k_1$ between 5 and 21, with all possible values of $k_2$ to $k_5$ considered. Chords with *GCD* > 1 were rejected. The remaining chords were sorted by increasing *CY* value. For a given *CY*, many reorderings of the same chord were found. The first entries for the first ten *CY* values are given in the following table.

**Table 21: Pentatonic scale examples**

| $k_1$ | $k_2$ | $k_3$ | $k_4$ | $k_5$ | $2k_1$ | ***OddComplexity OCY*** | ***Complexity CY*** | ***MinRatio MR*** | ***MinRatio*** (cents) |
|---|---|---|---|---|---|---|---|---|---|
| 8 | 9 | 10 | 12 | 15 | 16 | 45 | 720 | 16/15 | 111.73 |
| 12 | 14 | 15 | 20 | 21 | 24 | 105 | 840 | 21/20 | 84.47 |
| 12 | 14 | 16 | 18 | 21 | 24 | 63 | 1008 | 9/8 | 203.91 |
| 15 | 18 | 20 | 24 | 27 | 30 | 135 | 1080 | 10/9 | 182.40 |
| 15 | 16 | 20 | 24 | 25 | 30 | 75 | 1200 | 25/24 | 70.67 |
| 9 | 10 | 12 | 14 | 15 | 18 | 315 | 1260 | 15/14 | 119.44 |
| 20 | 22 | 24 | 30 | 33 | 40 | 165 | 1320 | 12/11 | 150.64 |
| 14 | 18 | 21 | 24 | 27 | 28 | 189 | 1512 | 28/27 | 62.96 |
| 20 | 24 | 26 | 30 | 39 | 40 | 195 | 1560 | 40/39 | 43.83 |
| 8 | 10 | 12 | 14 | 15 | 16 | 105 | 1680 | 16/15 | 111.73 |

These are just a selection of pentatonic scales available in Just Intonation. However, there aren't any scales with *Complexity* less than 720. If searches were done for longer scales, say 7 or 12 note scales, the minimal *CY* value increases rapidly. Some examples found include: (32, 36, 40, 45, 48, 54, 60, 64) has CY = 8640; (360, 400, 405, 432, 450, 480, 486, 540, 576, 600, 648, 675, 720) has CY = 388 800.

One interesting feature in Table 21 is minimum ratio between consecutive notes, *MinRatio*. Looking at the last column, most of the minimum ratios for pentatonic scales are below 120 cents, meaning they have narrow intervals. Two exceptions include (12, 14, 16, 18, 21, 24) and (15, 18, 20, 24, 27, 30) with minimum ratio 204 cents and 182 cents respectively. The first pentatonic scale has seventh harmonics, the second has fifth harmonics. An interesting feature is that the first scale outperforms the second in terms of *CY*, *OCY* and *MinRatio*, even though it uses seventh harmonics instead of fifth harmonics. Working with integers generally produces interesting and exceptional cases, and justly intoned mathematical harmony is no exception.